\documentstyle[12pt,epsfig]{article}
\makeatletter
\def\slash#1{{\mathpalette\c@ncel{#1}}} 
\makeatother

\newcommand\beq{\begin{eqnarray}}
\newcommand\eeq{\end{eqnarray}}

\def\xhat{\widehat{x}}
\def\zhat{\widehat{z}}

\begin{document}
\begin{flushright}
\end{flushright}
\vspace*{15mm}
\begin{center}
{\Large \bf 
Single Transverse-Spin Asymmetry in
Large $P_T$\\[2mm] 
Open Charm Production
at an Electron-Ion Collider}
\vspace{1.5cm}\\
 {\sc Hiroo Beppu$^1$, Yuji Koike$^2$, Kazuhiro Tanaka$^3$ and Shinsuke Yoshida$^1$}
\\[0.4cm]
\vspace*{0.1cm}{\it $^1$ Graduate School of Science and Technology, Niigata University,
Ikarashi, Niigata 950-2181, Japan}\\
\vspace*{0.1cm}{\it $^2$ Department of Physics, Niigata University,
Ikarashi, Niigata 950-2181, Japan}\\
\vspace*{0.1cm}{\it $^3$ Department of Physics, 
Juntendo University, Inzai, Chiba 270-1695, Japan}
\\[3cm]



{\large \bf Abstract} \end{center}
We discuss the single transverse-spin asymmetry (SSA)
to be observed in the $D$-meson production with large transverse-momentum
in semi-inclusive deep inelastic scattering, $e p^\uparrow  \rightarrow  e D X$.
This contribution is embodied as a twist-3 mechanism in the collinear factorization,
which is induced by purely gluonic correlation inside the transversely-polarized nucleon,
in particular, by the three-gluon correlation effects.
The complete formula for the corresponding SSA in the leading-order QCD
is expressed in terms of the four independent gluonic correlation functions
and reveals the five independent structures with respect to the dependence
on the azimuthal angle for the produced $D$-meson.
We present the numerical calculations of the SSA formula 
at the kinematics relevant to a future Electron Ion Collider.

\newpage
\section{Introduction}
Charm productions in semi-inclusive deep inelastic scattering (SIDIS), 
as well as in the $pp$ collision, are known to
be associated with the twist-2 gluon distributions in the nucleon, since
the $c\bar{c}$-pair creation through the photon-gluon or gluon-gluon fusion
is their driving subprocess. 
Similarly, the twist-3 contributions in the charm productions can be generated by
the purely gluonic effects inside the nucleon, in particular, the multi-gluon correlations.
Indeed, the observation of 
the single transverse-spin asymmetry (SSA) in the open charm productions 
allows us to probe the corresponding 
twist-3 effects~\cite{Kang:2008qh,Kang:2008ih,Beppu:2010qn,Koike:2011ns,Koike:2010jz}.

The corresponding SSA
arises as a {\it naively T-odd} effect 
in the cross section for the scattering of transversely-polarized nucleon 
off an unpolarized particle,
observing a 
$D$-meson 
with momentum $P_h$ 
in the final state, and this requires,
(i) nonzero transverse-momentum $P_{h\perp}$ originating 
from transverse motion
of 
quark or gluon; 
(ii) nucleon helicity flip in the cut diagrams for the cross section, corresponding
to the transverse polarization; 
and (iii) interaction beyond Born level to produce the
interfering phase between the LHS and the RHS of the cut in those diagrams. 
In particular, 
for large $P_{h\perp}\gg \Lambda_{\rm QCD}$,
the contribution~(i) arises perturbatively
as the recoil from the hard (unobserved) final-state partons, 
and this leads us to the collinear-factorization framework,
so that the other two contributions, (ii) and (iii), 
are generated by the twist-3 mechanism
associated with the three-gluon correlation functions 
for the transversely-polarized nucleon~\cite{Beppu:2010qn}.
This twist-3 mechanism
may be considered as an extension of the corresponding mechanism 
for
the SSA in the pion productions in the SIDIS~\cite{EKT07}, $pp$ collisions~\cite{Qi,To}, etc., 
based on the quark-gluon correlations in the nucleon, but
it has been clarified~\cite{Beppu:2010qn} that a straightforward extension~\cite{Kang:2008qh,Kang:2008ih} 
to the $D$-meson productions leads to missing many terms in the SSA.
The complete leading-order (LO) QCD formulae for the corresponding SSA
in the $D$-meson productions have been recently
derived in~\cite{Beppu:2010qn} for SIDIS and in \cite{Koike:2010jz} for $pp$ collisions,
based on the relevant twist-3 mechanism, i.e., the soft-gluon-pole 
mechanism,
and those results revealed, for the first time,
the entire nonperturbative gluonic degrees of freedom to induce the SSA
and the whole structures of the asymmetries with respect to 
the azimuthal angle for the produced $D$-meson.
Remarkably, those complete twist-3 formulae are related to 
the certain derivative of the twist-2 cross sections for the $D$-meson
productions~\cite{Koike:2011ns,Koike:2010jz},
thanks to universal structure behind the SSAs 
in a variety of hard processes~\cite{KT071,KT072}.

The purpose of this paper
is to present a numerical estimate of the SSA 
in the high-$P_{h\perp}$ $D$-meson production
in SIDIS, $ep^\uparrow \rightarrow eDX$,
based on the LO QCD formula of \cite{Beppu:2010qn}.
We use the modeling~\cite{Koike:2010jz} of the twist-3 
three-gluon correlation functions in the nucleon,
which is guided by the SSA for the $D$-meson production 
observed at RHIC~\cite{Liu:2009zzw},
and demonstrate the influence of the nonperturbative behaviors
of gluonic correlations.
We calculate the SSA 
at the kinematics 
relevant to Electron Ion Collider (EIC)~\cite{Anselmino:2011ay}.

\section{The LO QCD formula for the $D$-meson production}
To describe the $D$-meson production
in SIDIS, $e(\ell)+p(p,S_\perp)\to e(\ell')+D(P_h)+X$, 
we use, as usual, the kinematic variables $S_{ep}=(\ell+p)^2$, $q=\ell-\ell'$, $Q^2=-q^2$, $x_{bj}=Q^2/(2p\cdot q)$, 
and $z_f=p\cdot P_h/ (p\cdot q)$.
We work in a frame where the 3-momenta $\vec{q}$ and $\vec{p}$ are collinear, 
both moving along the $z$-axis as $q^\mu =g^{\mu 3}Q$ and $p^\mu=g^{\mu -}Q/(\sqrt{2}x_{bj})$, 
and define $q_T \equiv P_{h\perp}/z_f$ and the azimuthal angles around the $z$-axis $\phi$, $\Phi_S$, and $\chi$
of the lepton plane, the spin vector $S_\perp^\mu$, and the $D$-meson momentum $P_h^\mu$, 
respectively~\cite{Beppu:2010qn,Koike:2011ns}.
We take into account the masses $m_c$ and $m_h$ for the charm quark and the $D$ meson.

\begin{figure}
\includegraphics[height=3.6cm]{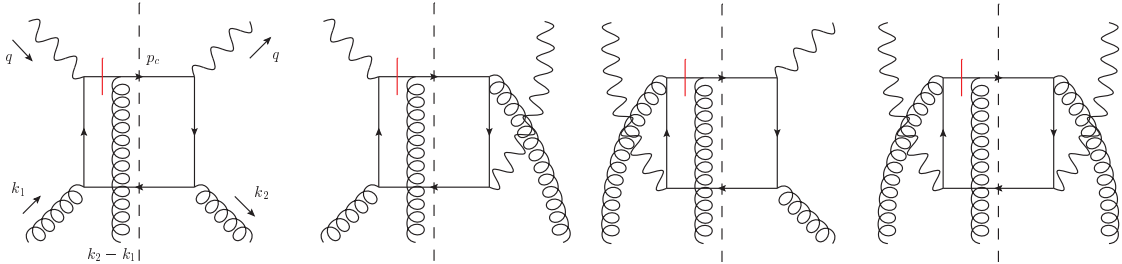}
\caption{Feynman diagrams for partonic subprocess in $ep^\uparrow\to eDX$;
mirror diagrams also contribute.}
\label{fig:1}
\end{figure}
In the LO in QCD perturbation theory, 
the photon-gluon fusion subprocesses of Fig.~1 
drive the SSA for large $P_{h\perp}$ $D$-meson production;
in Fig.~1,
the above-mentioned contribution~(i) is provided by the recoil from the hard unobserved $\bar c$ quark
and the $c$ quark with the momentum $p_c$ fragments into the $D$-meson in the final state.
The short bar on the internal $c$-quark line
indicates that the pole part is to be taken from the corresponding propagator,
to produce the
interfering phase for the contribution~(iii);
we note that these pole contributions from Fig.~1 would cancel the similar contributions 
from the corresponding mirror diagrams, if the $c$ quark were unobserved in
the final state as in the case of the $\bar D$-meson production.
The external curly lines represent the gluons that are generated from the three-gluon
correlations present inside the transversely-polarized nucleon, 
$\langle p S_\perp| A_{\alpha}(0)A_{\beta}(\eta) A_{\gamma}(\xi)|p S_\perp \rangle$, corresponding to 
the contribution~(ii). 
The diagrams obtained by the permutation of the gluon lines in Fig.~1 also produce the 
contributions~(i)-(iii), but
the Bose statistics of the gluons in the above matrix element 
guarantees that we need not consider those diagrams separately.
Thus, the SSA in the present context 
can be derived entirely as the contributions of soft-gluon-pole (SGP) type~\cite{EKT07}, 
leading to $k_2-k_1=0$, 
by evaluating the pole part in Fig.~1.~\footnote{The other types of pole contributions
participate in the SSA in other processes, 
see \cite{EKT07,To,Kanazawa:2011er}.}
The twist-3 nature of those contributions are unraveled
by the collinear expansion, as usual.
The expansion produces lots of terms, 
each of which is not gauge invariant.
Indeed, many of them vanish or cancel eventually, and the remaining terms can be 
organized into a gauge-invariant form.
This can be demonstrated~\cite{Beppu:2010qn} by 
sophisticated use of the Ward identities
for the contributions of the diagrams in Fig.~1.
The resulting factorization formula of the spin-dependent, differential cross section
for $ep^\uparrow\to eDX$ reads~\cite{Beppu:2010qn,Koike:2011ns}
\beq
&&
\hspace{-0.7cm}
\frac{d^6\Delta\sigma
}{[d\omega]}
=\frac{\alpha_{em}^2\alpha_se_c^2 M_N}{16\pi^2
 x_{bj}^2S_{ep}^2Q^2}\! \left(\frac{-\pi}{2}\right) \!
\sum_{k=1,\cdots,4,8,9}
{\cal A}_k{\cal S}_k 
\nonumber\\
&&\times
\int_{x_{\rm min}}^1
\frac{dx}{x}\!
\int_{z_{\rm min}}^1
\frac{dz}{z} 
\delta \! \left(
\frac{q_T^2}{Q^2}-\! \left(1-\frac{1}{\hat{x}}\right) \! \left(1-\frac{1}{\hat{z}}\right)\!
+\frac{m_c^2}{\hat{z}^2Q^2}\right)
\nonumber\\
&&
\!\!\!\!\!\!
\times\! 
\sum_{a=c,\bar{c}}
D_a(z) 
\left(\!\delta_a \!\left\{
\left[\frac{dO(x,x)}{dx}-\frac{2O(x,x)}{x}\right]\! \Delta\hat{\sigma}^{1}_k
+\!\left[\frac{dO(x,0)}{dx}-\frac{2O(x,0)}{x}\right]\! \Delta\hat{\sigma}_k^2
\right. \right.
\nonumber\\
&&
\!\!\!\!\!\!\!\!\!\!\!\!
\left.\left.\left. \left.
+\!\frac{O(x,x)}{x}\Delta\hat{\sigma}^{3}_k
+\!\frac{O(x,0)}{x}\Delta\hat{\sigma}^{4}_k
\right\} 
+\right\{ O(x,x)\rightarrow N(x,x),\;  O(x,0)\rightarrow - N(x,0) \right\} \right)\! ,
\label{3gluonresult}
\eeq
where $[d\omega] \equiv dx_{bj}dQ^2dz_fdq_T^2d\phi d\chi$ denotes 
the differential elements,
$\hat{x}=x_{bj}/x$ and $\hat{z}=z_f/z$
are the partonic variables associated with the usual
momentum fractions $x$ and $z$, respectively,
$D_c(z)$ denotes the 
usual twist-2 fragmentation function for a $c$-quark to become the $D$-meson, and
the quark-flavor index $a$ can, in principle, be $c$ and $\bar{c}$, with $\delta_c=1$ and  
$\delta_{\bar{c}}=-1$, so that the cross section for 
the $\bar{D}$-meson production $ep^\uparrow\to e\bar{D}X$
can be obtained by 
a simple replacement of the fragmentation function 
to that for the $\bar{D}$ meson, $D_a (z) \rightarrow \bar{D}_a (z)$. 
$O(x_1, x_2)$ and $N(x_1, x_2)$ represent a complete set of twist-3 
gluonic correlation functions,
defined through
the gauge-invariant lightcone correlation~\footnote{We suppress the gauge-link 
operators to be inserted in between the field strength tensors.}
of three field-strength tensors~\cite{Beppu:2010qn},
\begin{eqnarray}
&&
\hspace{-0.7cm}
{\cal M}^{\alpha\beta\gamma}_{F(-)}(x_1,x_2)\! \equiv \!
-gi^3\!\!\int{d\lambda\over 2\pi}\int{d\zeta\over 2\pi}e^{i\lambda x_1}
e^{i\zeta(x_2-x_1)}\langle p S_\perp|d^{bca}F_b^{\beta n}(0)F_c^{\gamma n}(\zeta n)F_a^{\alpha n}(\lambda n)
|p S_\perp\rangle
\nonumber\\
&&
\!\!
=\! 2iM_N \!\!
\left[
O(x_1,x_2)g^{\alpha\beta}\epsilon^{\gamma pnS_\perp}
\!+\! O(-x_2,x_1-x_2)g^{\beta\gamma}\epsilon^{\alpha pnS_\perp}
\!+\! O( x_2-x_1, -x_1)g^{\gamma\alpha}\epsilon^{\beta pnS_\perp}\right] ,
\label{3gluonO}\\
&&
\hspace{-0.7cm}
{\cal M}^{\alpha\beta\gamma}_{F(+)}(x_1,x_2)\! \equiv \!
-gi^3\!\!\int{d\lambda\over 2\pi}\int{d\zeta\over 2\pi}e^{i\lambda x_1}
e^{i\zeta(x_2-x_1)}\langle p S_\perp|if^{bca}F_b^{\beta n}(0)F_c^{\gamma n}(\zeta n)F_a^{\alpha n}(\lambda n)
|p S_\perp\rangle
\nonumber\\
&&
\!\!
=\! 2iM_N \!\!
\left[
N(x_1,x_2)g^{\alpha\beta}\epsilon^{\gamma pnS_\perp}
\!+\! N(-x_2,x_1-x_2)g^{\beta\gamma}\epsilon^{\alpha pnS_\perp}
\!+\! N( x_2-x_1, -x_1)g^{\gamma\alpha}\epsilon^{\beta pnS_\perp}\right] ,
\label{3gluonN}
\end{eqnarray}
with the nucleon mass $M_N$ and a lightlike vector $n$ satisfying $n^2=0$ and $p\cdot n=1$;
$d^{bca}$ and $f^{bca}$ are, respectively, 
the symmetric and anti-symmetric structure constants of the color SU(3) group,
so that $O(x_1, x_2)$ and $N(x_1, x_2)$ are the $C$-odd and $C$-even functions
satisfying 
\begin{eqnarray}
&&O(x_1, x_2)=O(x_2, x_1)=O(-x_1, -x_2) ,
\nonumber\\
&&N(x_1, x_2)=N(x_2, x_1)=-N(-x_1, -x_2) .
\label{sym}
\end{eqnarray}
In (\ref{3gluonresult}), $\alpha_{em}$
is the fine-structure constant,
$e_c=2/3$ is the electric charge of the $c$-quark, and 
the summation for the subscript $k$ runs over $k=1,2,3,4,8,9$, with
\begin{eqnarray}
&&{\cal A}_1=1+\cosh^2\psi,
\;\;\;\;\;\;\;\;
{\cal A}_2=-2,
\;\;\;\;\;\;\;\;
{\cal A}_3=-\cos(\phi-\chi)\sinh 2\psi, 
\nonumber\\
&&{\cal A}_4=\cos 2(\phi-\chi)\sinh^2\psi,
\;\;\;\;\;
{\cal A}_8=-\sin(\phi-\chi)\sinh 2\psi, 
\;\;\;\;\;
{\cal A}_9=\sin 2(\phi-\chi)\sinh^2\psi,
\nonumber\\
&&{\cal S}_1={\cal S}_2={\cal S}_3={\cal S}_4=\sin(\Phi_S-\chi) , 
\;\;\;\;\;\;\;\;
{\cal S}_8={\cal S}_9=\cos(\Phi_S-\chi) ,
\label{a19}
\end{eqnarray}
where $\cosh\psi \equiv 2x_{bj}S_{ep}/ Q^2 -1$.
The delta function in (\ref{3gluonresult}) implies that
the lower limits of the integrals are given by~\cite{Kang:2008qh,Beppu:2010qn}
\begin{equation}
z_{\rm min}=z_f\frac{\left( 1 - x_{bj} \right)  Q^2}{2 x_{bj}m_c^2}  \left( 1
         - \sqrt{1- \frac{4x_{bj}m_c^2}{\left( 1 - x_{bj} \right) Q^2}
             \left[ 1
                + \frac{x_{bj} q_T^2}{\left( 1 - x_{bj} \right)  Q^2} \right]}\right) , 
\label{zmin}
\end{equation}
and
\begin{equation}
x_{\rm min} = \left\{ 
\begin{array}{ll}
x_{bj}\left[1+\frac{z_f^2q_T^2+m_c^2}{z_f(1-z_f)Q^2}\right]
& \quad \mbox{for}\;\;\; z_f\left(1+\sqrt{1+\frac{q_T^2}{m_c^2}}\right) > 1,\\ 
\\
x_{bj} \left[1+\frac{2m_c^2}{Q^2}\left(1+\sqrt{1+\frac{q_T^2}	 
{m_c^2}}\right)\right]
& \quad \mbox{for}\;\;\; z_f\left(1+\sqrt{1+\frac{q_T^2}{m_c^2}}\right) \leq 1.\\
\end{array} \right. 
\label{xmin}
\end{equation}
Partonic hard parts $\Delta\hat{\sigma}_k^i$ 
depend on $m_c$ as well as other partonic variables;
for the explicit formulae of $\Delta\hat{\sigma}_k^i$, we refer the readers to 
Eqs.~(71)-(74) in \cite{Beppu:2010qn}.
The participation of the ``derivative terms'', the terms with the derivatives of the three-gluon 
correlation functions as $dO(x,x)/dx$, $dO(x,0)/dx$, $dN(x,x)/dx$, and $dN(x,0)/dx$,
is characteristic of the contributions originating from SGPs.
Note that, instead of evaluating the SGP contributions arising in the diagrams in Fig.~1 
as above,
those results can be obtained
using the ``master formula''~\cite{Koike:2011ns}, 
which is 
schematically given by
\begin{eqnarray}
\frac{d^6\Delta\sigma
}{[d\omega]}
\sim -i\pi
\sum_{a=c,\bar{c}}
\int\frac{dz}{z} D_a(z)
&& \!\!\!
\int
\frac{dx}{x^2}
{\partial {\cal H}_{\mu\nu}(xp,q, p_c)
\over \partial p_{c\perp}^\sigma}
\omega^\mu_{\ \alpha}\,\omega^\nu_{\ \beta}\,\omega^\sigma_{\ \gamma}\,
\nonumber\\
&&
\times
\left({\cal M}^{\alpha\beta\gamma}_{F(-)}(x,x) \delta_a   + {\cal M}^{\alpha\beta\gamma}_{F(+)} (x,x) 
\right) ,
\label{master}
\end{eqnarray}
where $\omega^\mu_{\ \,\nu}\equiv g^\mu_{\ \,\nu}-p^\mu n_\nu$, and 
${\cal H}_{\mu\nu}(xp,q, p_c)$ represents the
partonic hard part 
for the $2\to2$ Born subprocess, expressed by the diagrams in Fig.~1 with the
soft ($k_2-k_1=0$) gluon line removed.~\footnote{The color factors 
of the type, ${\rm Tr}[t^b t^c t^a ] = (d^{bca}+if^{bca})/4$, implied by the diagrams
in Fig.~1, are contracted with the three-gluon matrix elements,
giving rise to the combination of (\ref{3gluonO}) and (\ref{3gluonN}) in (\ref{master}).}
This reveals that $\Delta\hat{\sigma}_k^i$ in (\ref{3gluonresult}) are related 
to the twist-2 hard parts ${\cal H}_{\mu\nu}(xp,q, p_c)$,
similarly as 
in the SSA in various processes 
associated with 
twist-3 quark-gluon correlation functions~\cite{KT071,KT072}.

We now reexpress as 
\begin{equation}
\phi-\chi = \phi_h , \;\;\;\;\;\;\; \;\;\;\;\;\;\;
\Phi_S-\chi=\phi_h-\phi_S ,
\end{equation}
in
(\ref{a19}),
where $\phi_h$ and $\phi_S$ represent the azimuthal angles of the hadron plane 
and 
the nucleon's spin vector $\vec{S}_{\perp}$, respectively, measured from the {\it lepton plane}.
Then, (\ref{3gluonresult}) can be expressed as 
\begin{equation}
\frac{d^6\Delta \sigma}{[d\omega]}
= \sin(\phi_h -\phi_S)\left({\cal F}_1
+{\cal F}_2\cos\phi_h
+{\cal F}_3\cos2\phi_h \right)
+\cos(\phi_h -\phi_S)\left({\cal F}_4\sin\phi_h
+{\cal F}_5\sin2\phi_h \right) ,
\label{azimuth2}
\end{equation}
with the corresponding structure functions ${\cal F}_1,{\cal F}_2,\ldots, {\cal F}_5$, 
exhibiting the five independent azimuthal dependences similarly as
in the twist-3 SSA
for $ep^\uparrow\to e\pi X$, generated from the quark-gluon correlation functions~\cite{KT071}.  
Thus, the complete LO QCD formulae~(\ref{3gluonresult}), (\ref{azimuth2}) 
for the high-$P_{h\perp}$ $D$-meson production in SIDIS are expressed in terms of the
four types of gluonic functions $O(x,x)$, $O(x,0)$, $N(x,x)$, and $N(x,0)$ 
of the relevant momentum fraction $x$,
and generate five independent structures 
about the dependence on the relevant azimuthal angles.
This is in contrast to the corresponding results in \cite{Kang:2008qh},
which were expressed by only two types of gluonic functions
and three independent azimuthal structures.

The formulae (\ref{3gluonresult}), (\ref{azimuth2}) for the single-spin-dependent cross section 
should be compared 
with the corresponding LO QCD formulae of the twist-2 
unpolarized cross section for the high-$P_{h\perp}$ $D$-meson production in SIDIS,
which is generated from
the usual unpolarized gluon-density distribution $G(x)$, as~\cite{Beppu:2010qn}
\begin{eqnarray}
\frac{d^6\sigma^{\rm unpol}}{[d\omega]}
&&=\frac{\alpha_{em}^2\alpha_se_c^2}{64\pi^2
 x_{bj}^2S_{ep}^2Q^2}
\sum_{k=1}^4{\cal A}_k 
\int_{x_{\rm min}}^1
\frac{dx}{x} \!
\int_{z_{\rm min}}^1 
\frac{dz}{z}\delta \! \left(\frac{q_T^2}{Q^2}-
\left(1-\frac{1}{\hat{x}}\right)  \left(1-\frac{1}{\hat{z}}\right)
+\frac{m_c^2}{\hat{z}^2Q^2}\right)
\nonumber\\
&&
\;\;\;\;\;\;\;\;\;\;\;\;
\;\;\;\;\;\;\;\;\;\;\;\;\;\;\;\;\;\;
\;\;\;\;\;\;
\times
\sum_{a=c,\bar{c}} \! D_a(z)G(x)\hat{\sigma}^{U}_k 
\nonumber\\
&&= \sigma_1^U
+\sigma_2^U\cos\phi_h
+\sigma_3^U\cos2\phi_h ,
\label{unpolresult}
\end{eqnarray}
where the partonic hard cross sections are given in Eq.~(81) of \cite{Beppu:2010qn} and obey
\beq
\Delta\hat{\sigma}_k^1={2q_T\xhat\over Q^2(1-\zhat)} \hat{\sigma}^U_k ,
\label{unlike}
\eeq
as implied by (\ref{master}).

\section{SSA in the $D$-meson production at EIC}
We evaluate the SSAs for the $D^0$ production, $ep^\uparrow\to eD^0 X$, 
based on the QCD factorization formula (\ref{3gluonresult}).
In particular, using (\ref{azimuth2}) and (\ref{unpolresult}),
we calculate the asymmetries,
\begin{equation}
\frac{{\cal F}_1}{\sigma_1^U}, \;\;\;\;\; \;\;\;\;\;\frac{{\cal F}_2}{2 \sigma_1^U}, \;\;\;\;\;\;\;\;\;\;
\frac{{\cal F}_3}{2 \sigma_1^U}, \;\;\;\;\;\;\;\;\;\;
\frac{{\cal F}_4}{2 \sigma_1^U}, \;\;\;\;\;\;\;\;\;\;
\frac{{\cal F}_5}{2 \sigma_1^U},
\label{ssas}
\end{equation}
to be observed at a future EIC.
Because the partonic hard parts in (\ref{3gluonresult}) 
are common for the $C$-even and $C$-odd three-gluon functions,
we show
the contributions to (\ref{ssas})
from the $C$-odd correlation functions
$O(x,x)$ and $O(x,0)$ in the following calculations.
For the first estimate presented in this paper,
we assume
the two types of functional forms of those correlations, corresponding to the different small-$x$ behavior,
\beq
&&{\rm Model\ 1}:\qquad O(x,x)=O(x,0)=0.004\,x\,G(x),
\label{case1}\\
&&{\rm Model\ 2}:\qquad O(x,x)=O(x,0)=0.001\,\sqrt{x}\,G(x), 
\label{case2}
\eeq
where 
the coefficients $0.004$ and $0.001$ 
are suggested in \cite{Koike:2010jz} 
by the comparison of the three-gluon contribution to
the SSA for $p^\uparrow p\rightarrow DX$ with the data
observed at RHIC~\cite{Liu:2009zzw}.~\footnote{$O(x,x)$ and $O(x,0)$ in (\ref{case1}) and (\ref{case2})
are twice as large as those in \cite{Koike:2010jz}, by
which we take into account the effect of $N(x,x)$ and $N(x,0)$
contributing constructively as $O(x,x)$ and $O(x,0)$.} 
We use the CTEQ6L gluon-density distribution~\cite{Pumplin:2002vw}
for $G(x)$ of (\ref{unpolresult}), (\ref{case1}) and (\ref{case2}), 
and KKKS08 fragmentation function~\cite{Kneesch:2007ey} for $D_a(z)$ of (\ref{3gluonresult})
and (\ref{unpolresult}).
For those nonperturbative functions, we associate the scale $\mu^2 = Q^2+m_c^2 +z_f^2q_T^2$,
and, for simplicity, 
we assume that the scale dependence of $O(x,x)$ and $O(x,0)$
is determined by that of $G(x)$ according to (\ref{case1}) and (\ref{case2}).
We use $m_c=1.5$~GeV for the charm quark mass.

\begin{figure}
\vspace*{-1cm}
\includegraphics[height=7cm]{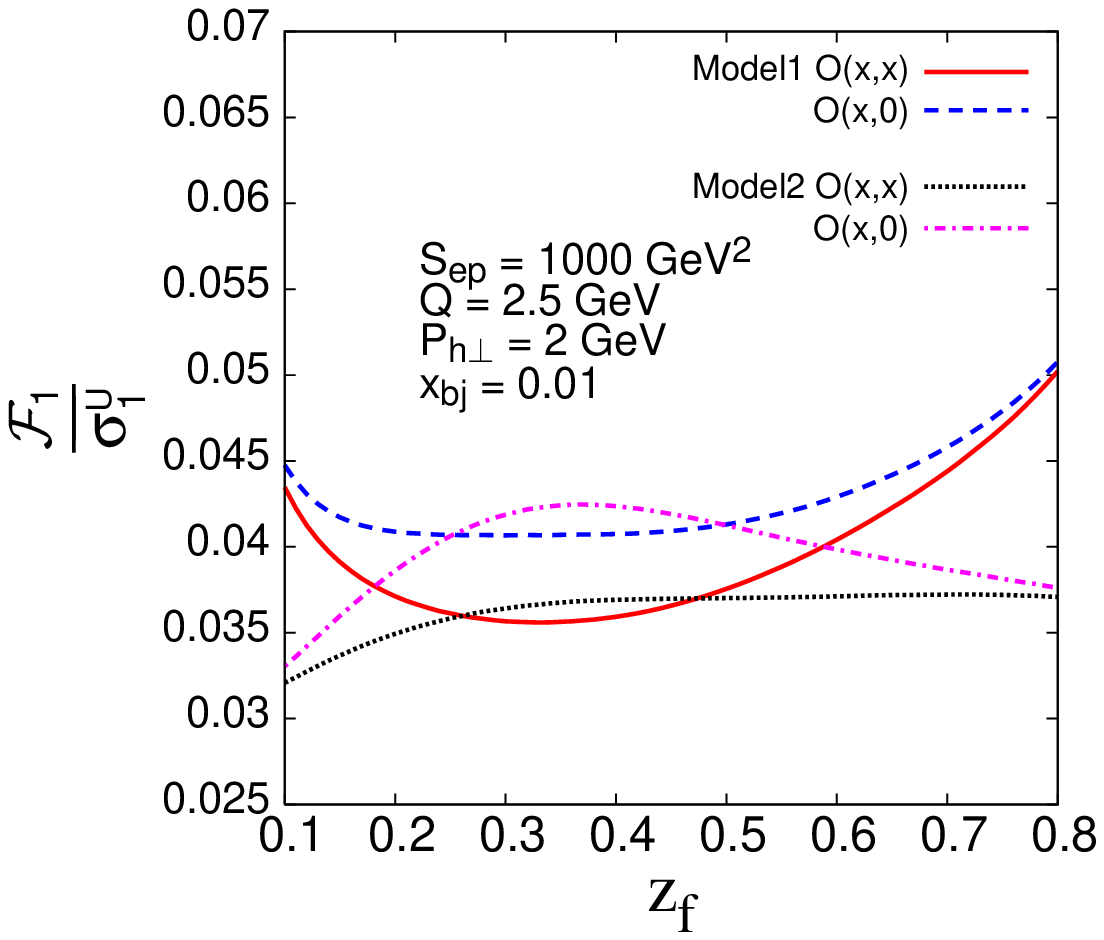}
\hspace{-0.4cm}
\includegraphics[height=7cm]{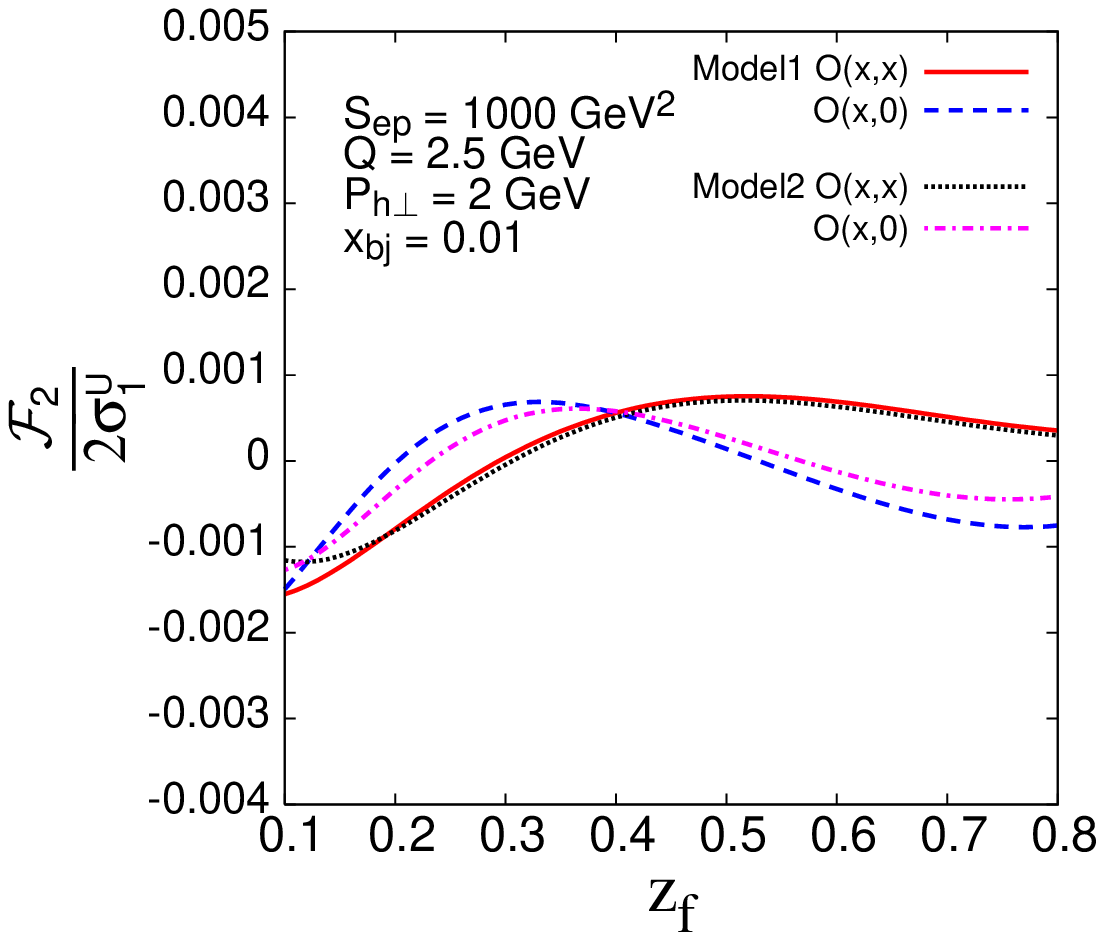}\\
\includegraphics[height=7cm]{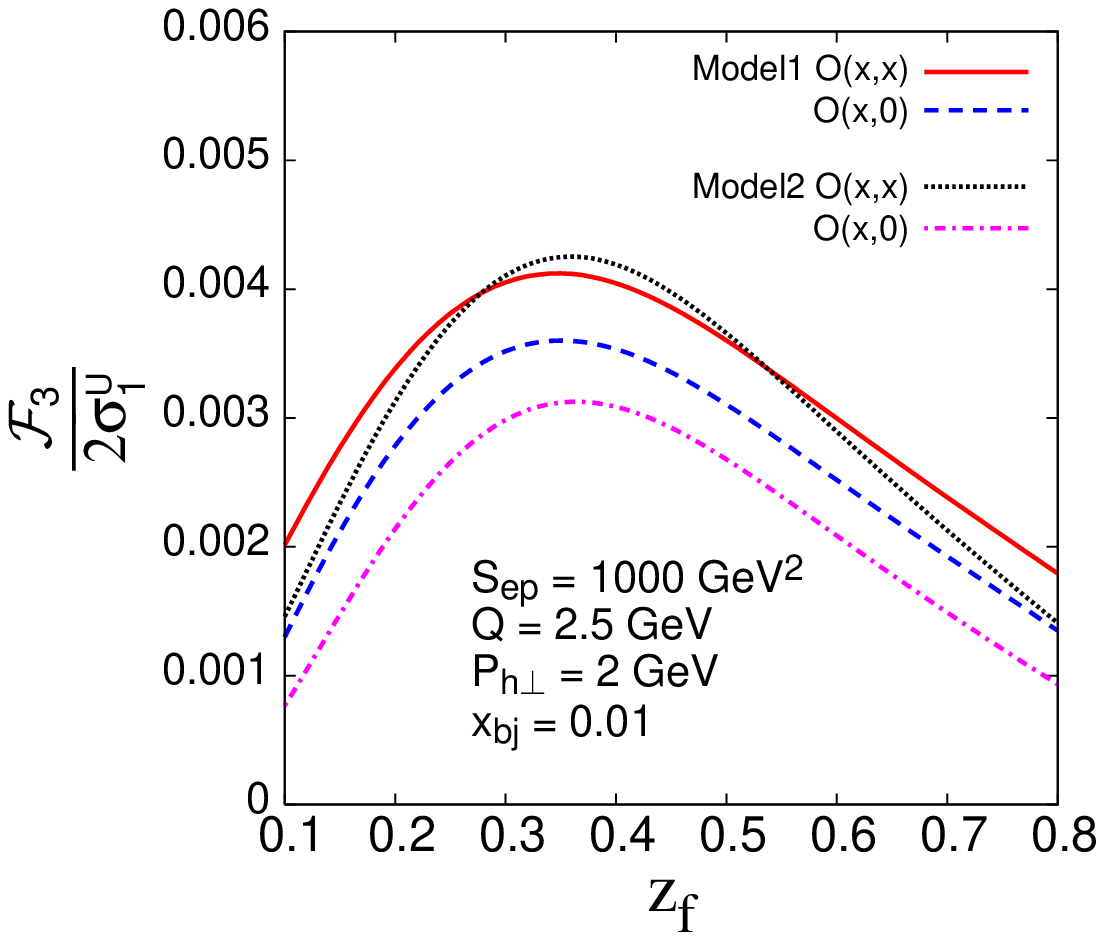}
\hspace{-0.4cm}
\includegraphics[height=7cm]{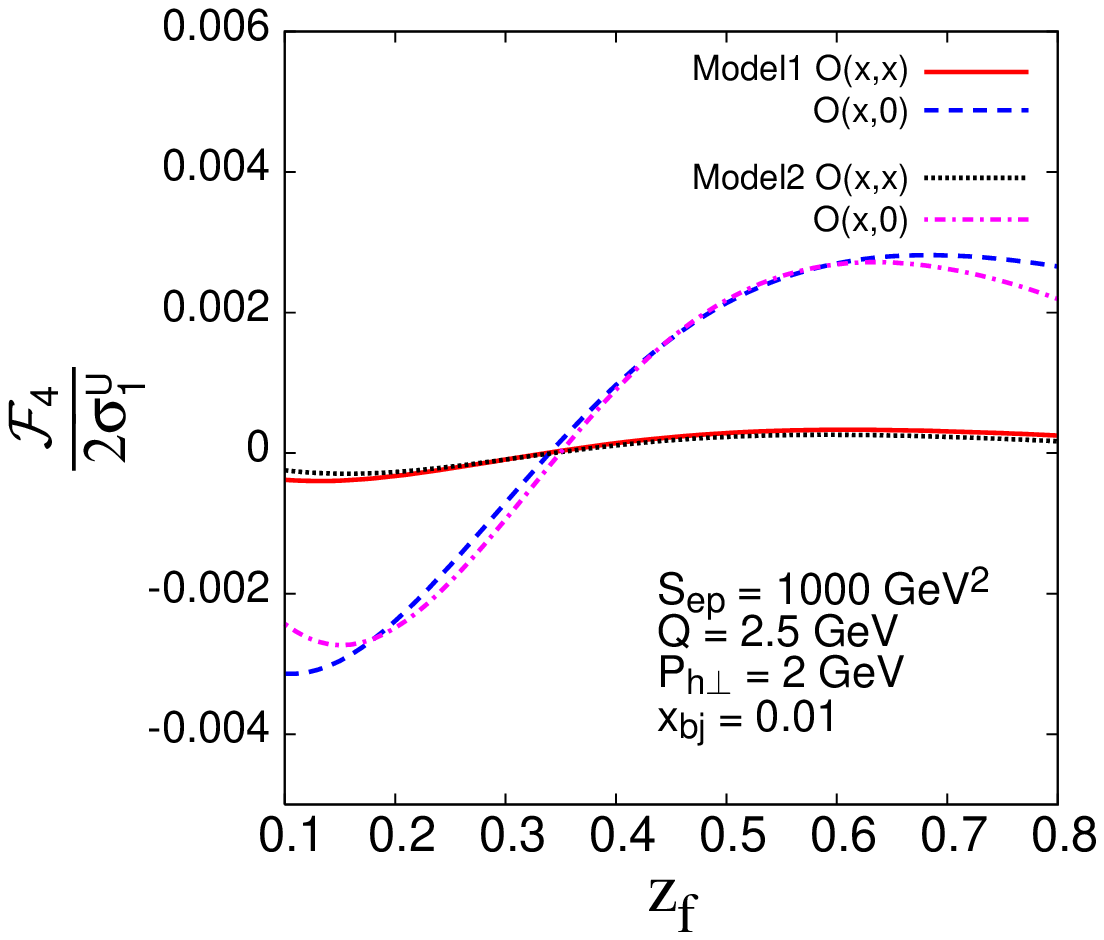}\\
\includegraphics[height=7cm]{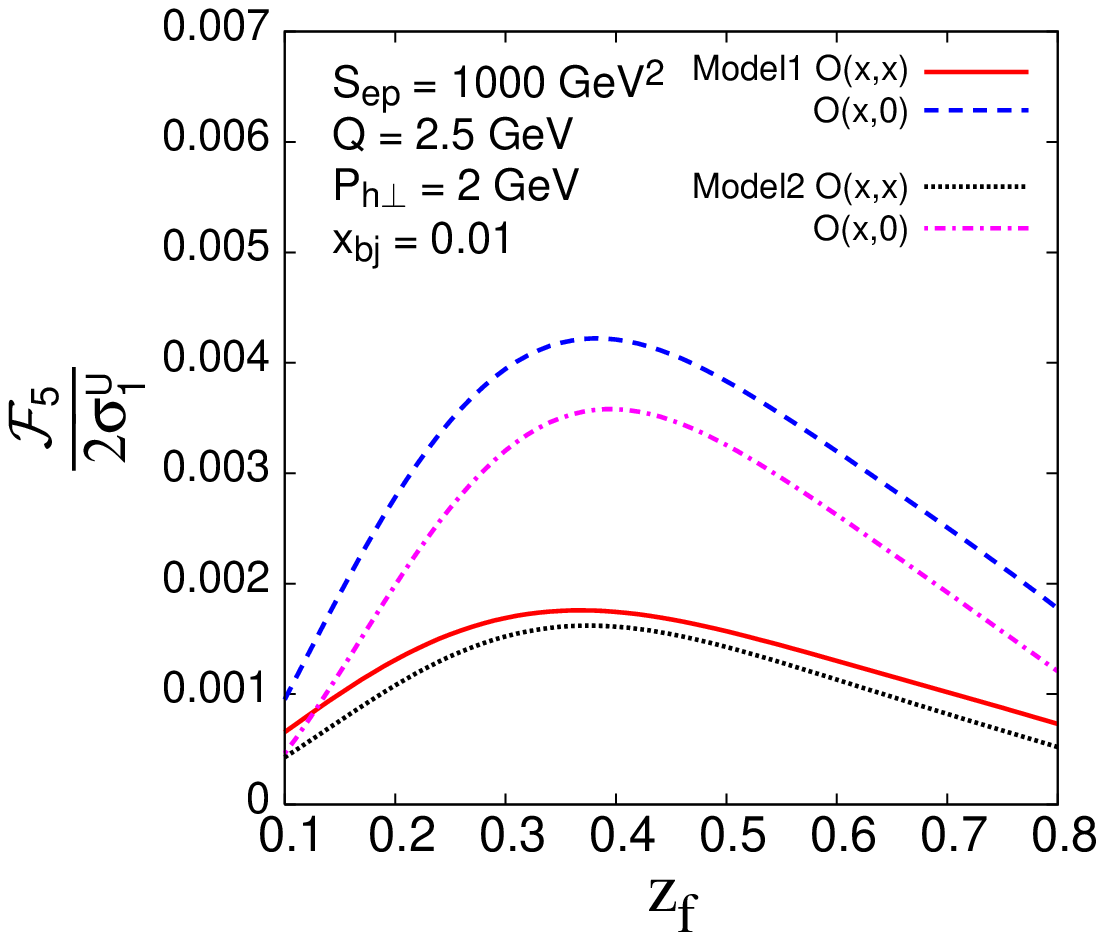}
\hspace{-0.4cm}
\includegraphics[height=7cm]{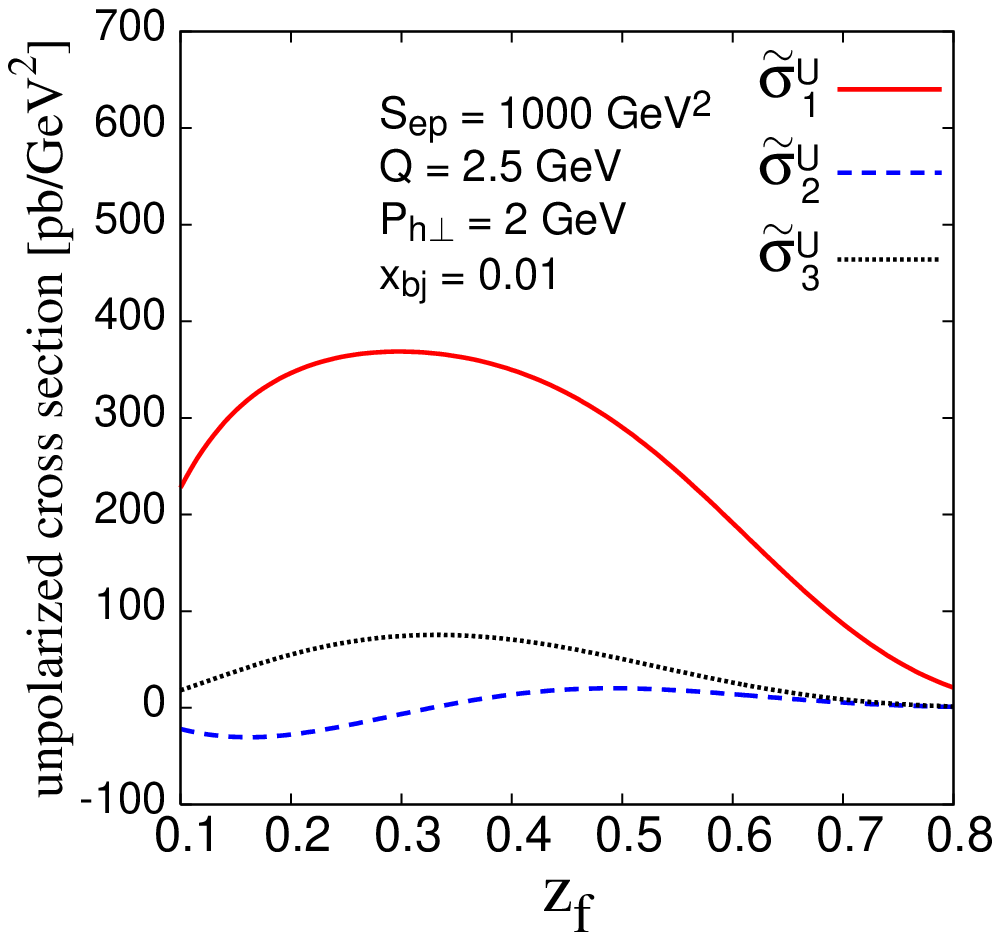}
\caption{The individual contributions of (\ref{case1}) and (\ref{case2})
to the SSAs~(\ref{ssas}) with (\ref{azimuth2}), (\ref{3gluonresult})
(the first five panels) and the individual coefficients of the unpolarized cross section~(\ref{unpolresult})
(the last panel),
plotted as a function of $z_f$,
for $D^0$ production in SIDIS 
at EIC kinematics
with $S_{ep}=1000$~GeV$^2$, $Q=2.5$~GeV, $P_{h\perp}=2$~GeV, and $x_{bj}=0.01$.}
\label{fig:21}
\end{figure}
\begin{figure}
\vspace*{-1.5cm}
\includegraphics[height=7cm]{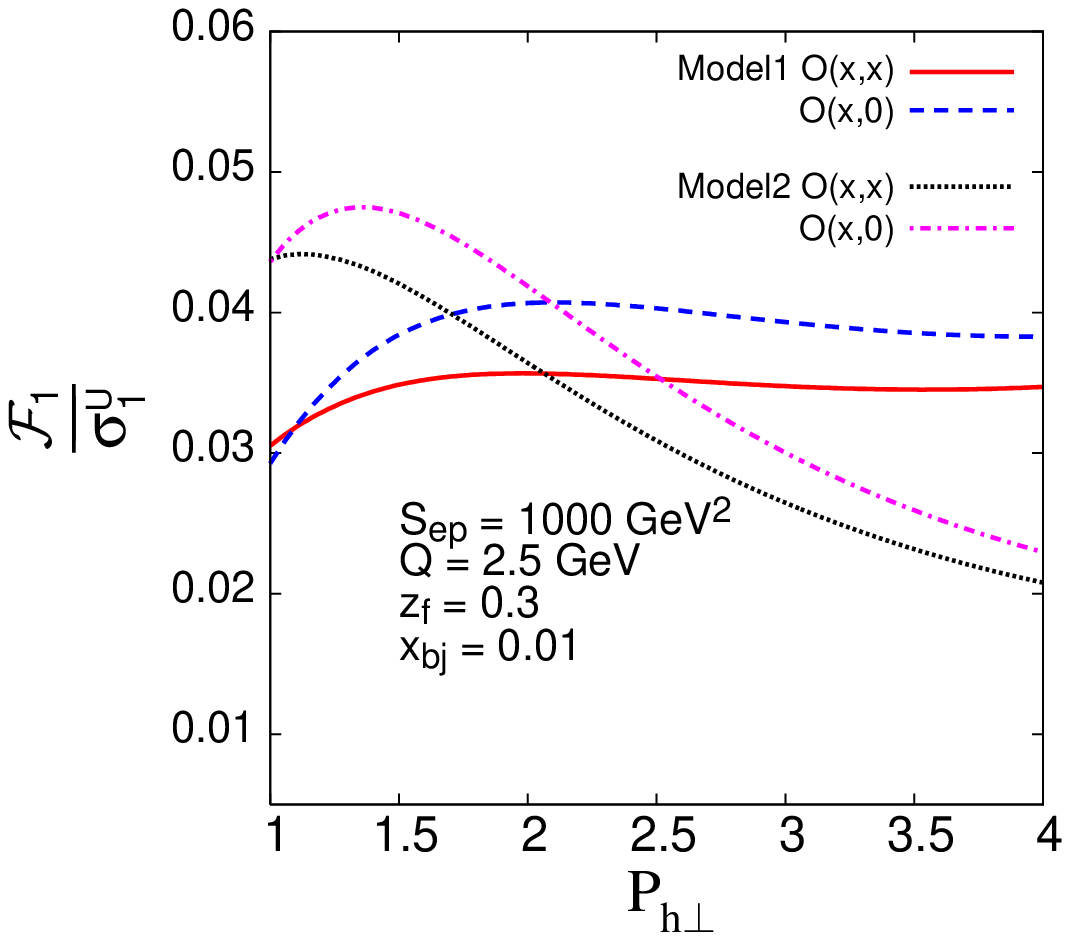}
\hspace{-0.4cm}
\includegraphics[height=7cm]{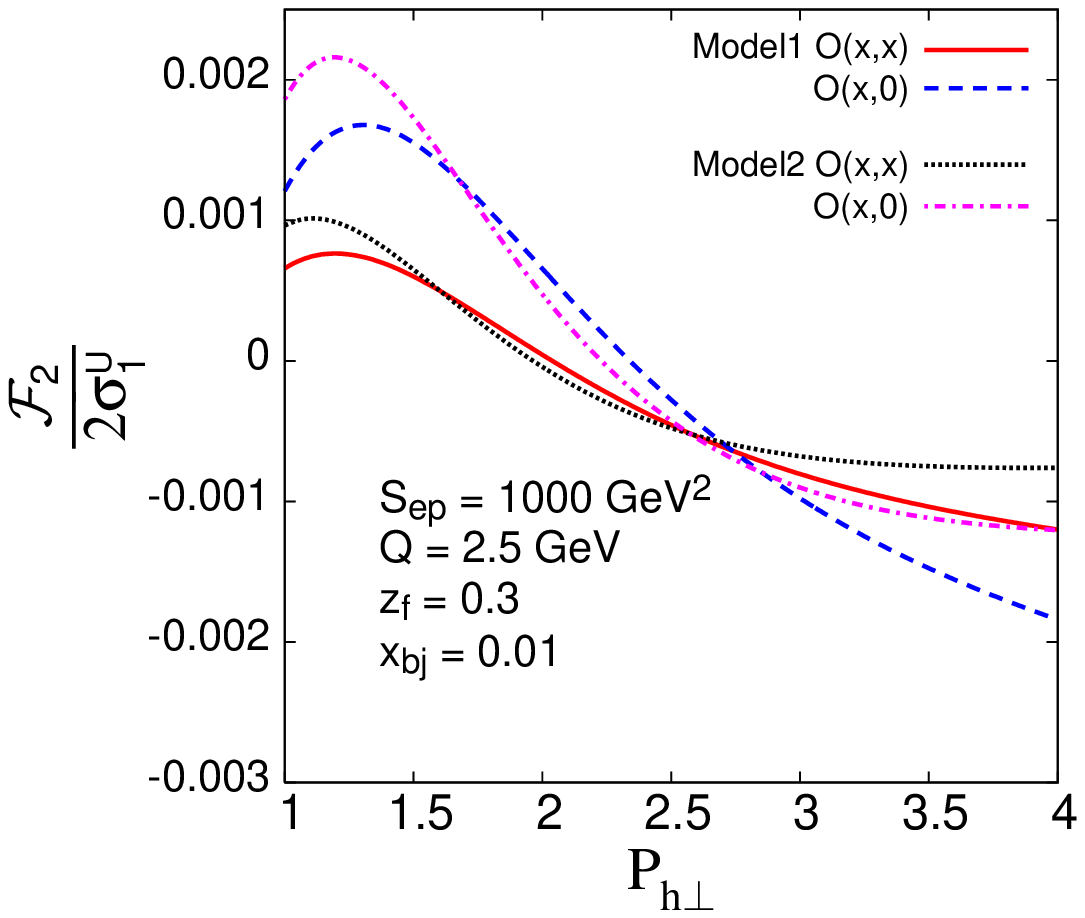}\\
\includegraphics[height=7cm]{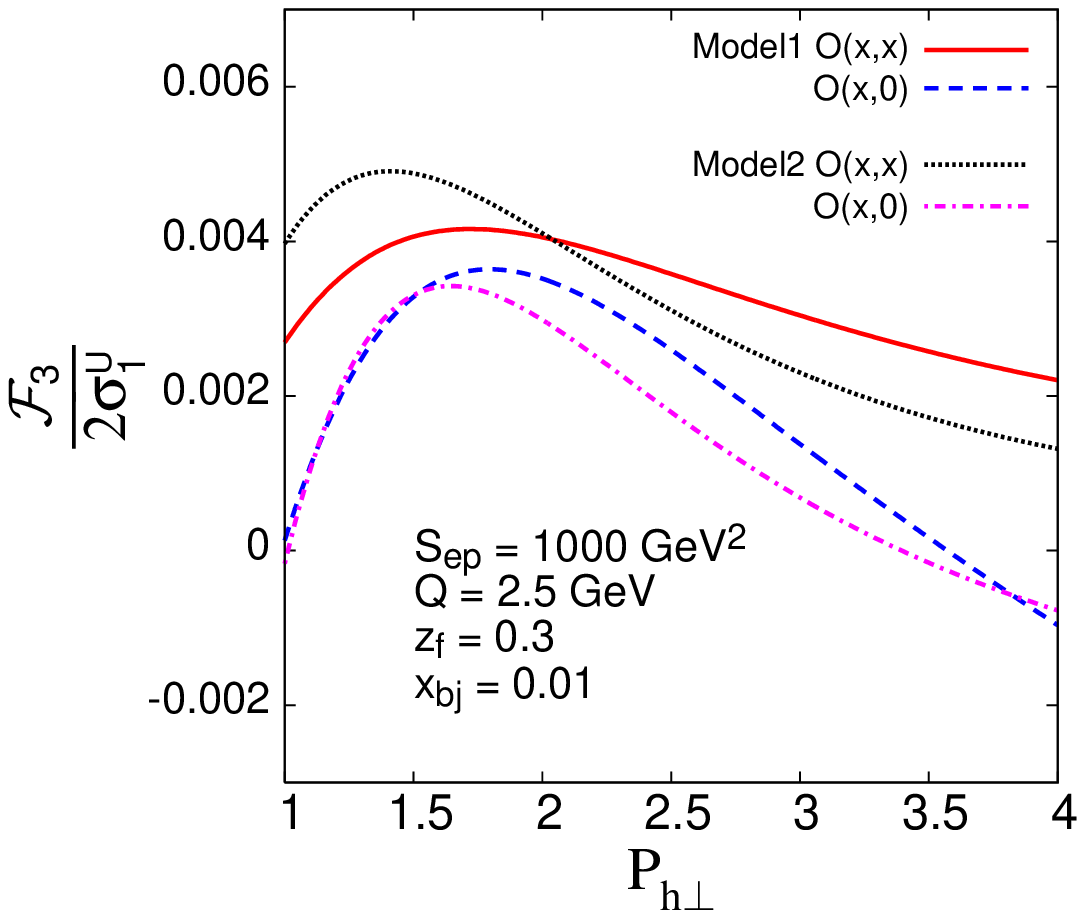}
\hspace{-0.4cm}
\includegraphics[height=7cm]{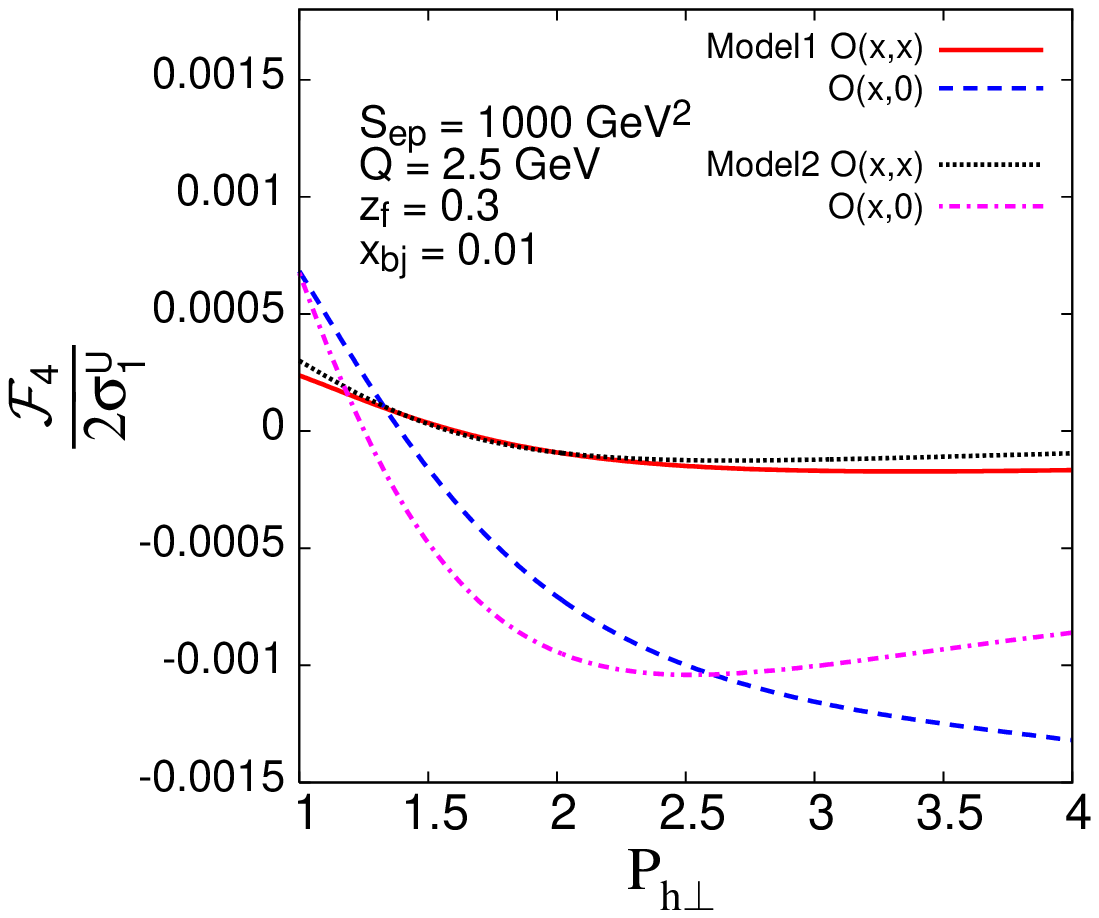}\\
\hspace{-0.4cm}
\includegraphics[height=7cm]{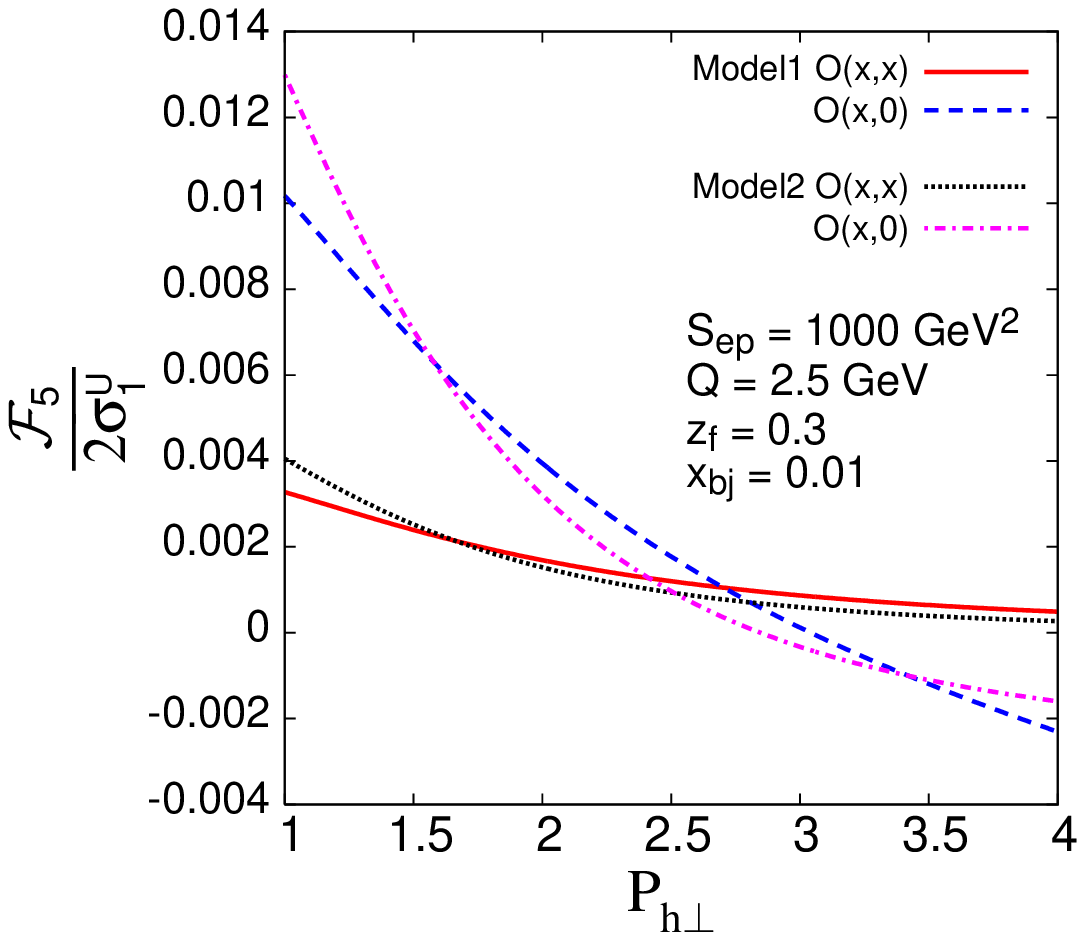}
\hspace{-0.4cm}
\includegraphics[height=7cm]{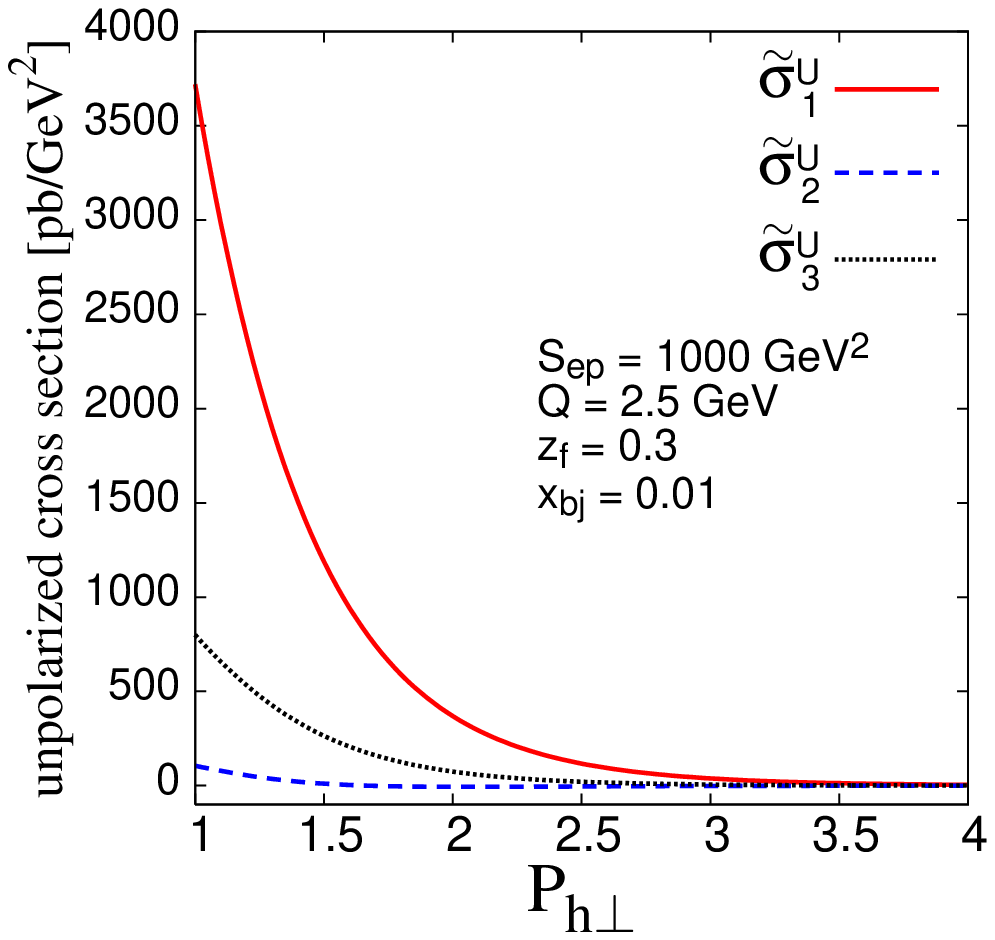}\\
\caption{The individual contributions of (\ref{case1}) and (\ref{case2})
to the SSAs~(\ref{ssas}) with (\ref{azimuth2}), (\ref{3gluonresult})
(the first five panels) and the individual coefficients of the unpolarized cross section~(\ref{unpolresult})
(the last panel),
plotted as a function of $P_{h\perp}$,
for $D^0$ production in SIDIS 
at EIC kinematics
with $S_{ep}=1000$~GeV$^2$, $Q=2.5$~GeV, $z_f=0.3$, and $x_{bj}=0.01$.}
\label{fig:22}
\end{figure}
\begin{figure}
\includegraphics[height=7cm]{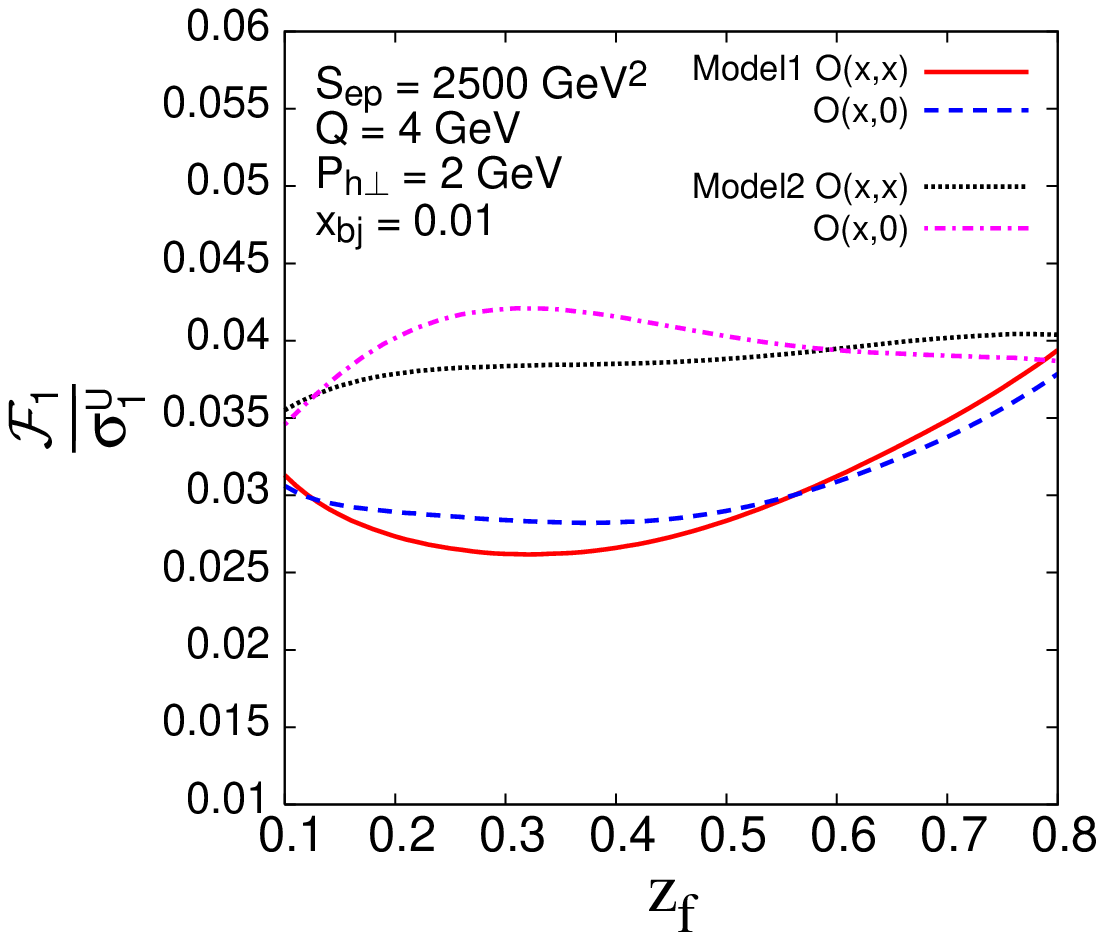}
\hspace{-0.4cm}
\includegraphics[height=7cm]{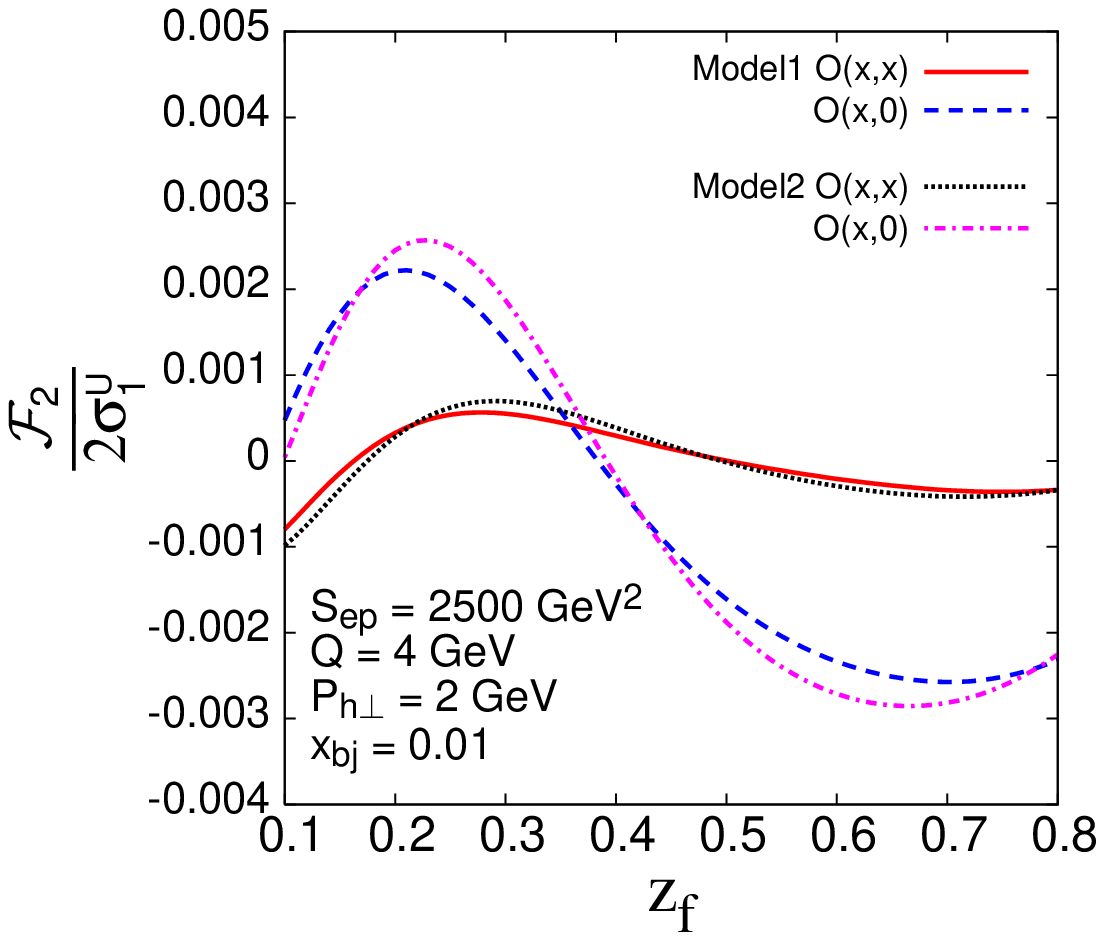}\\
\includegraphics[height=7cm]{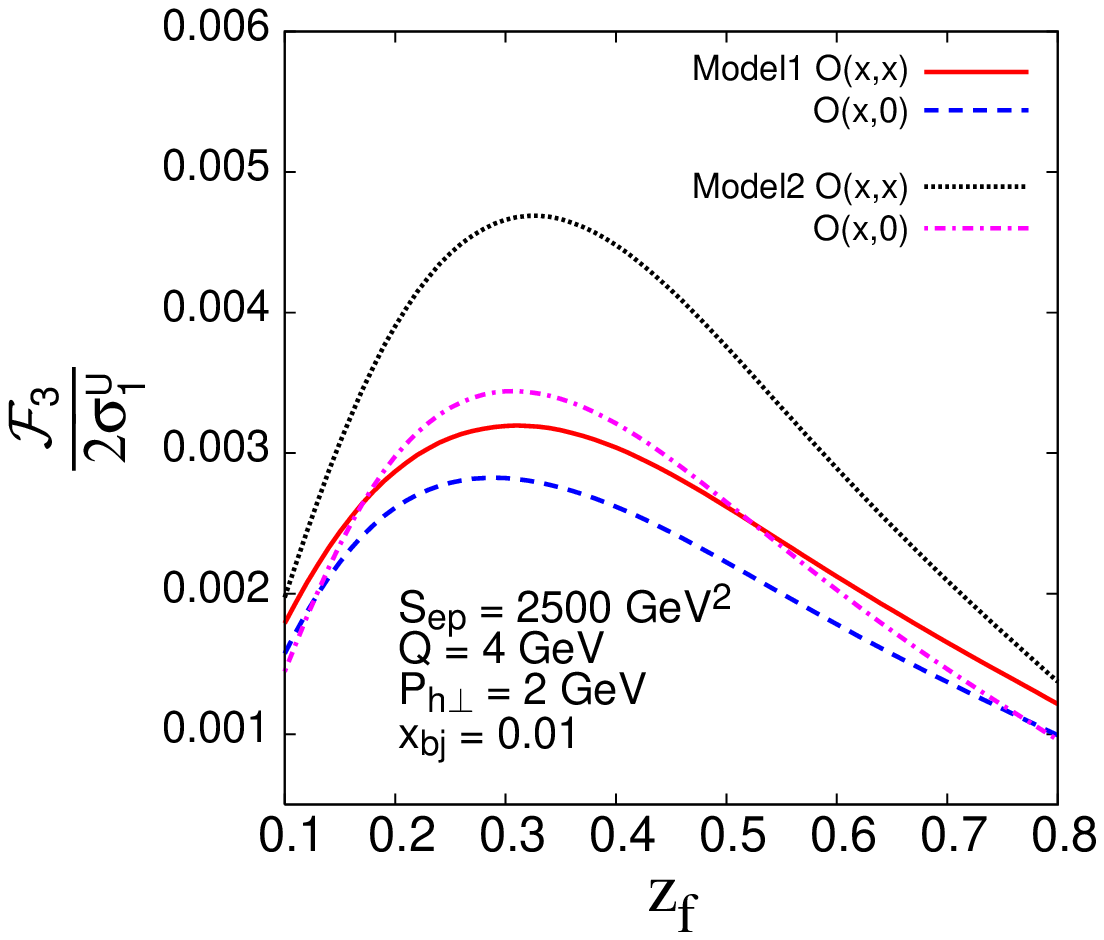}
\hspace{-0.4cm}
\includegraphics[height=7cm]{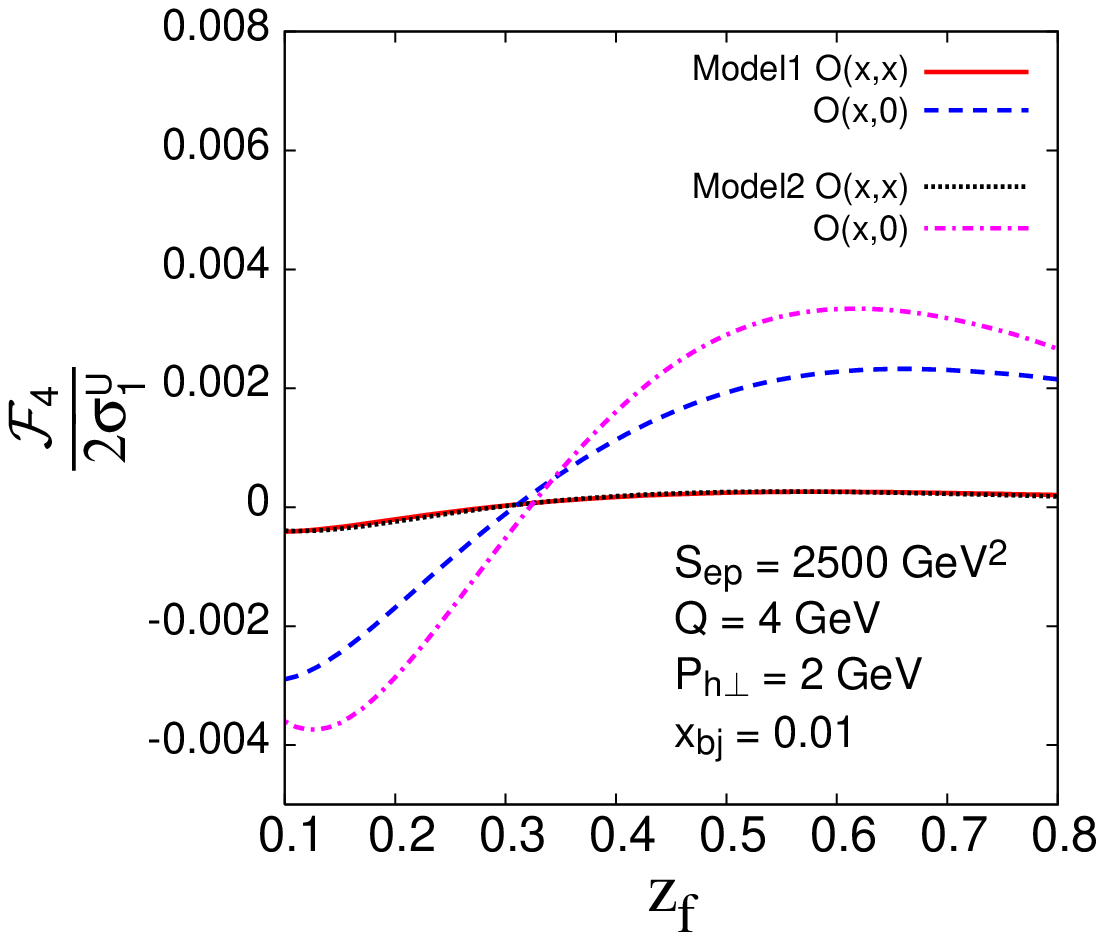}\\
\includegraphics[height=7cm]{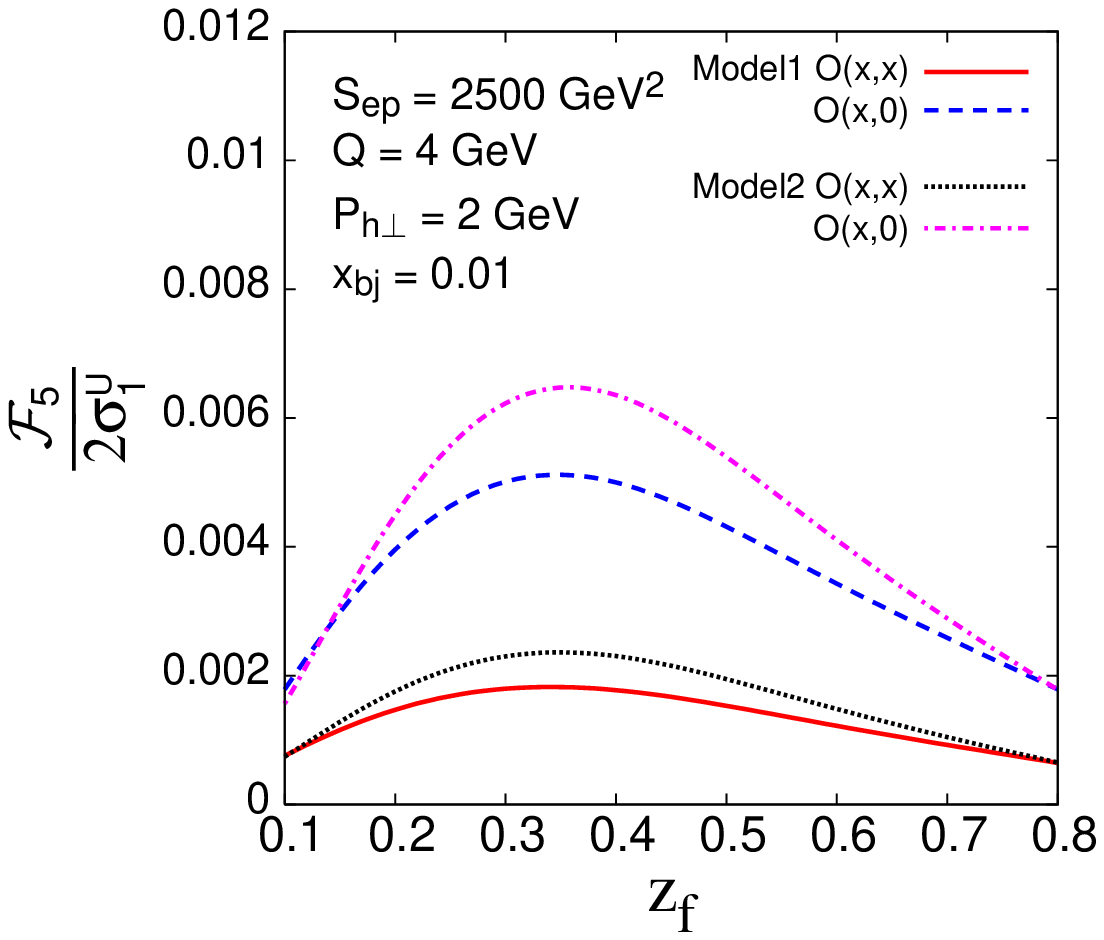}
\hspace{-0.4cm}
\includegraphics[height=7cm]{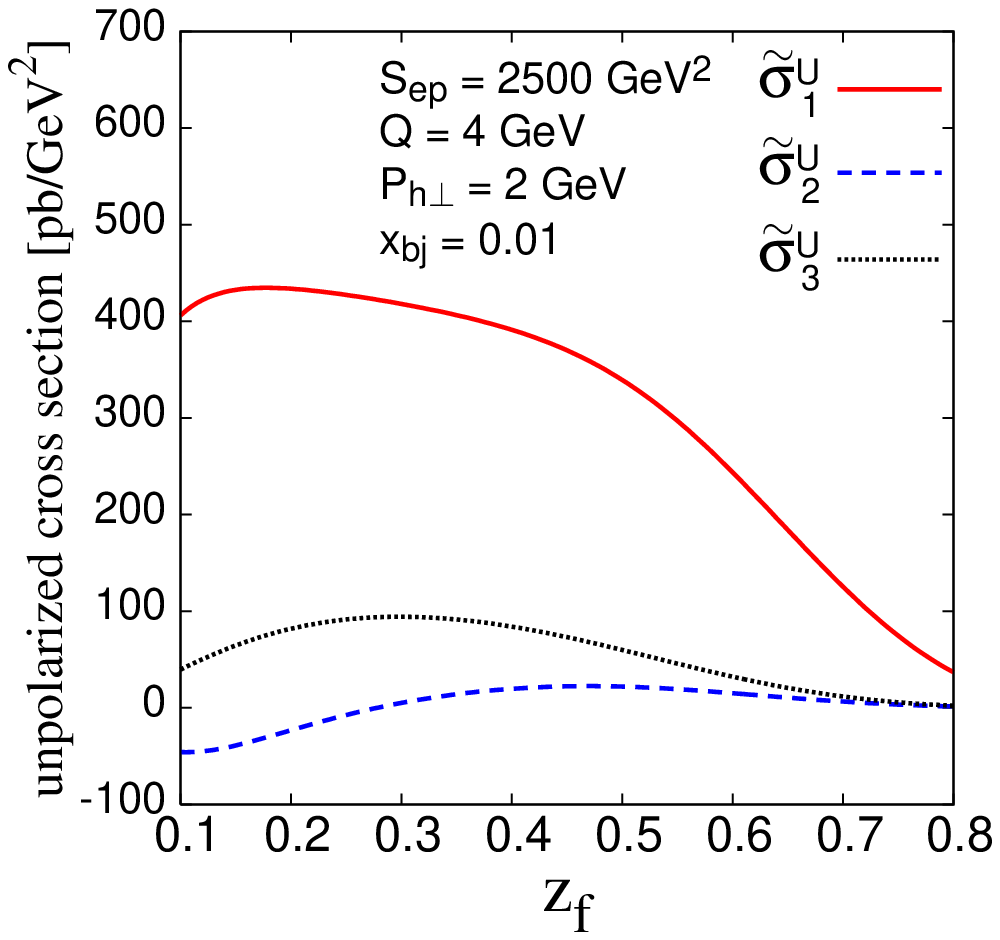}
\caption{Same as Fig.~\ref{fig:21}, but for 
EIC kinematics
with $S_{ep}=2500$~GeV$^2$, $Q=4$~GeV, $P_{h\perp}=2$~GeV, and $x_{bj}=0.01$.}
\label{fig:31}
\end{figure}
\begin{figure}
\includegraphics[height=7cm]{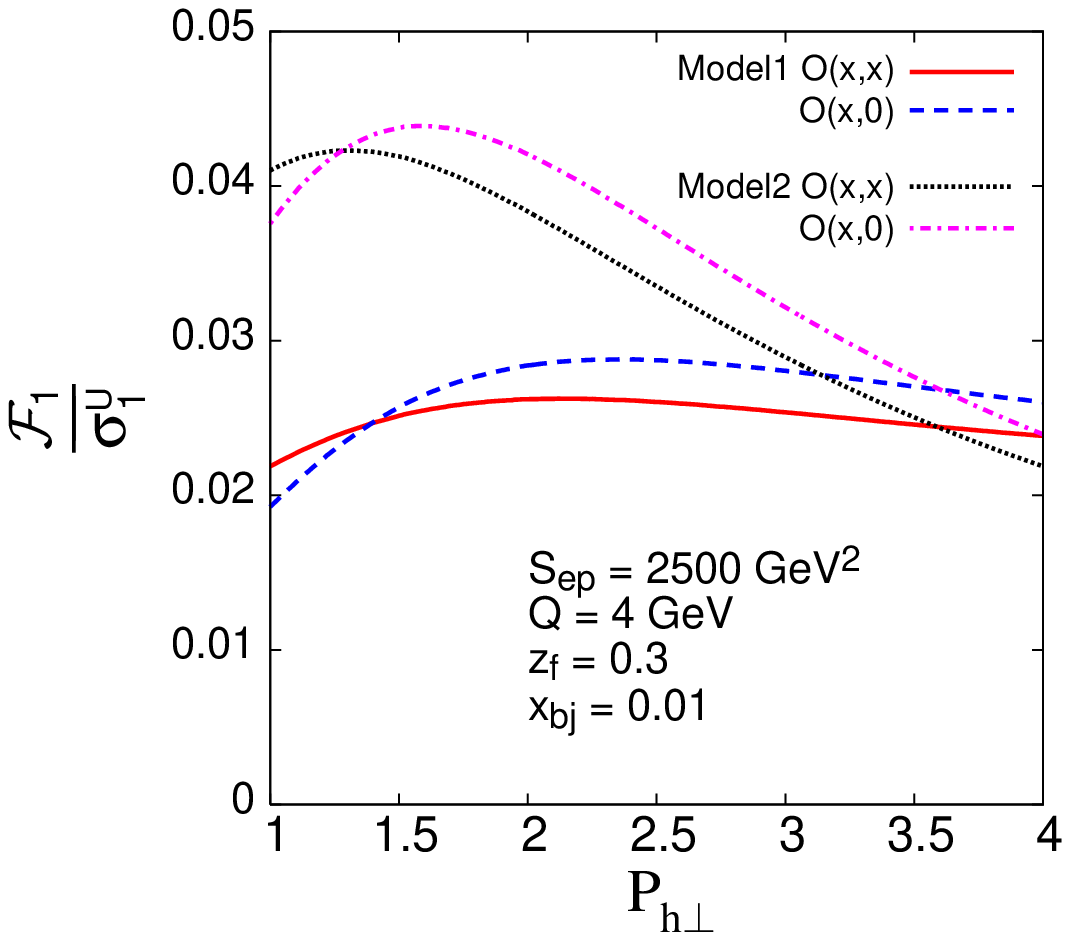}
\hspace{-0.4cm}
\includegraphics[height=7cm]{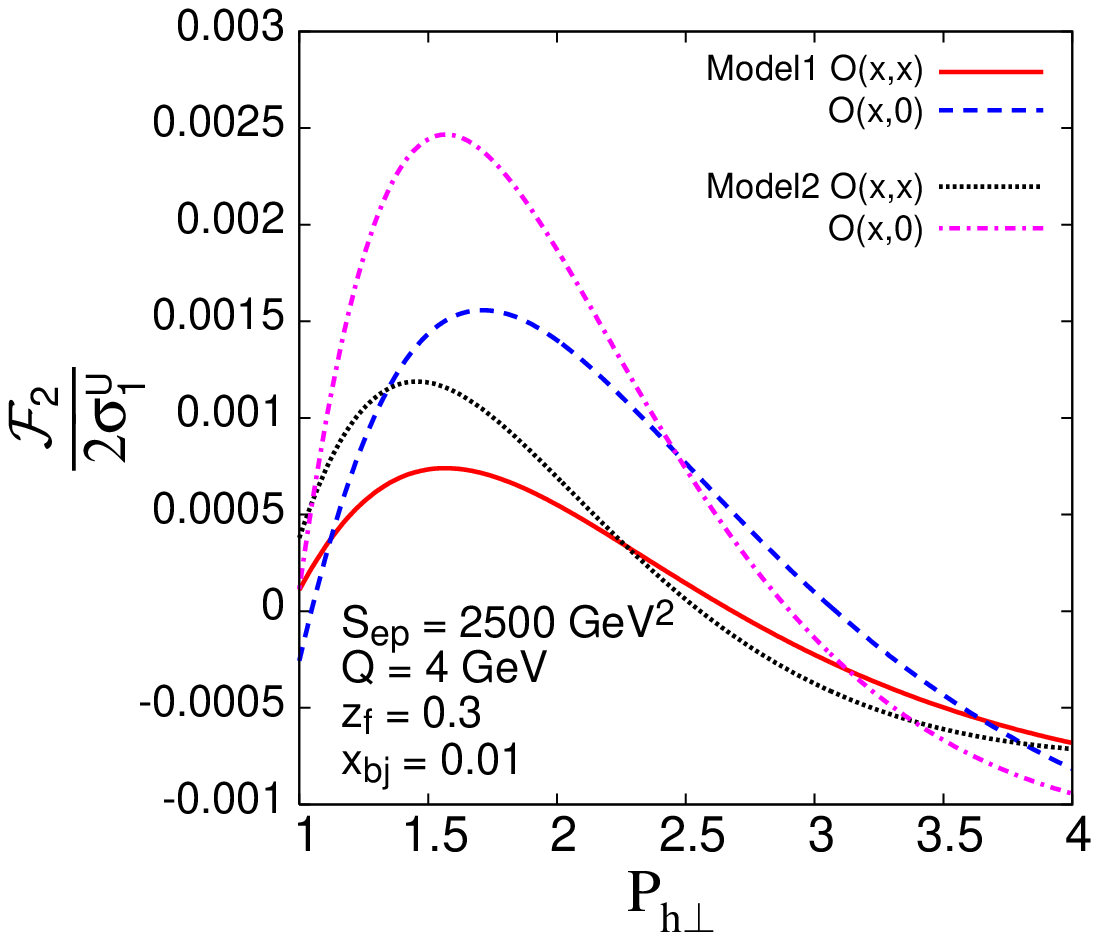}\\
\includegraphics[height=7cm]{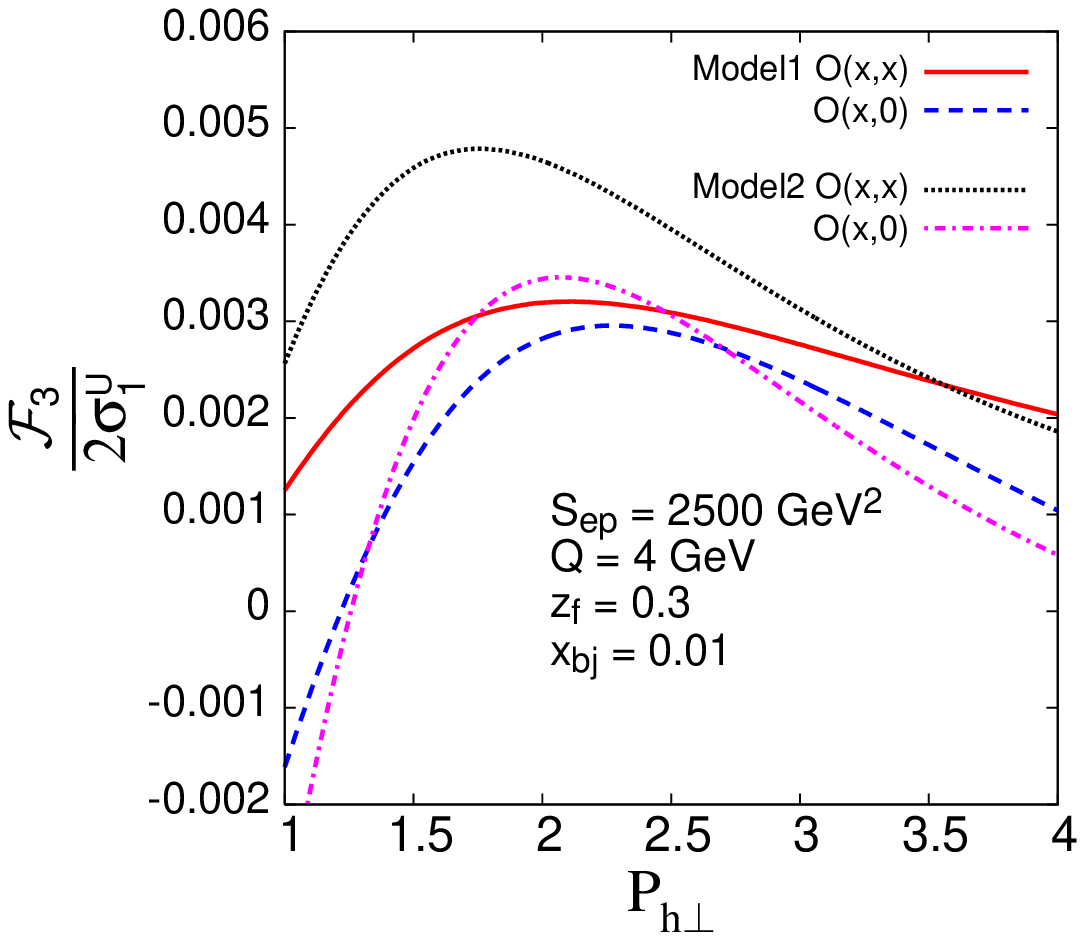}
\hspace{-0.4cm}
\includegraphics[height=7cm]{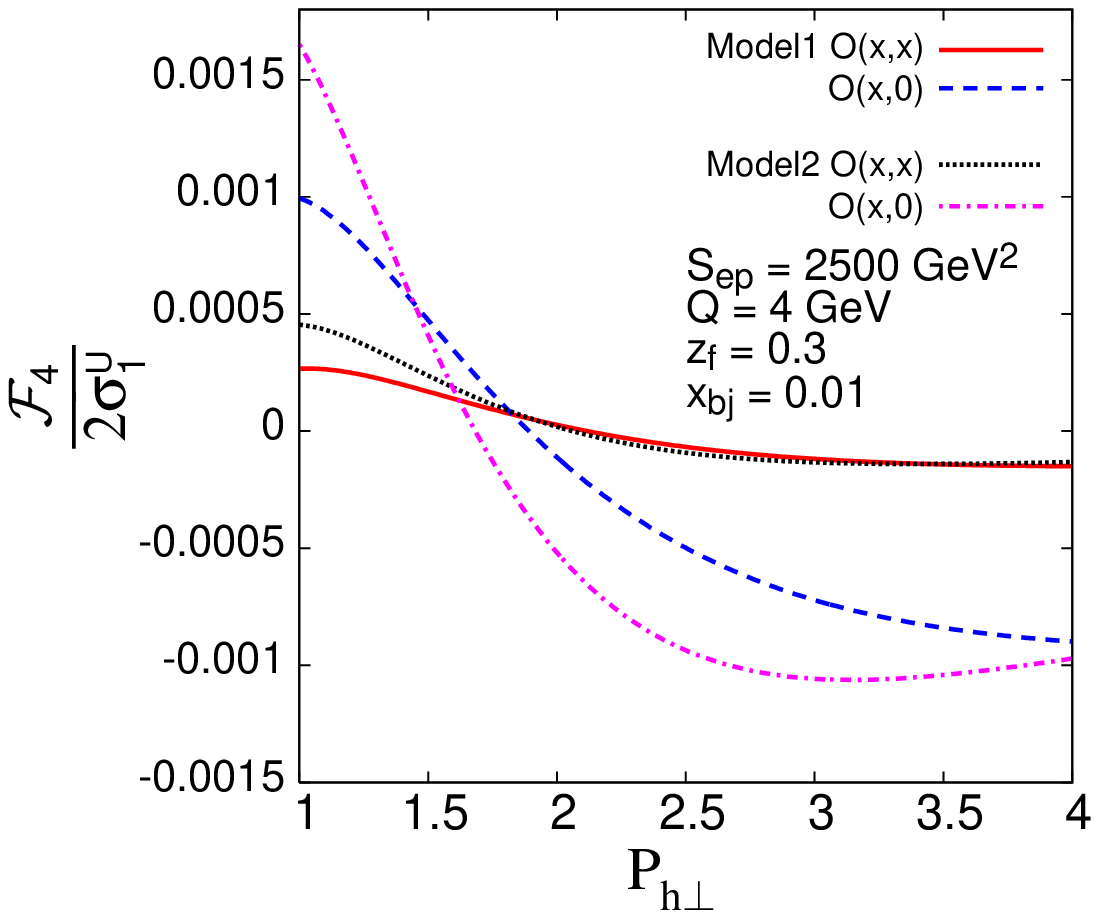}\\
\hspace{-0.4cm}
\includegraphics[height=7cm]{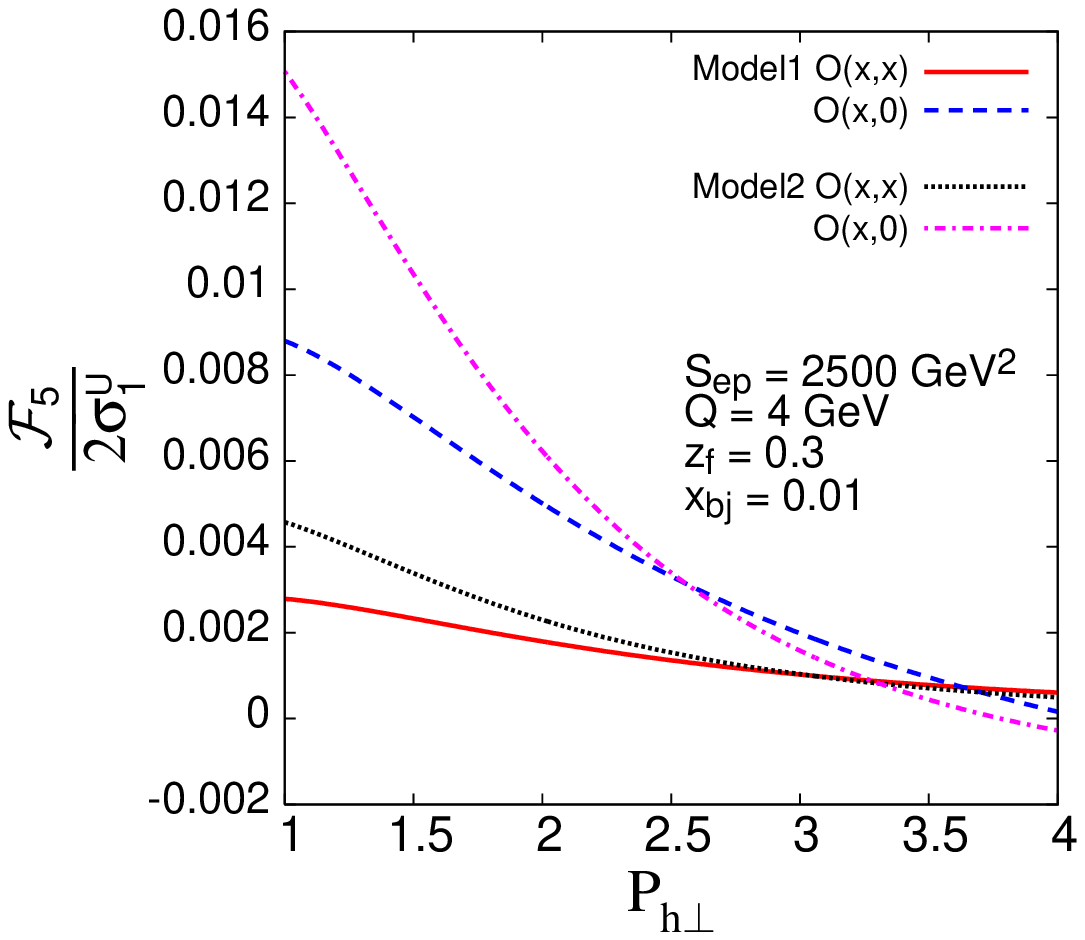}
\hspace{-0.4cm}
\includegraphics[height=7cm]{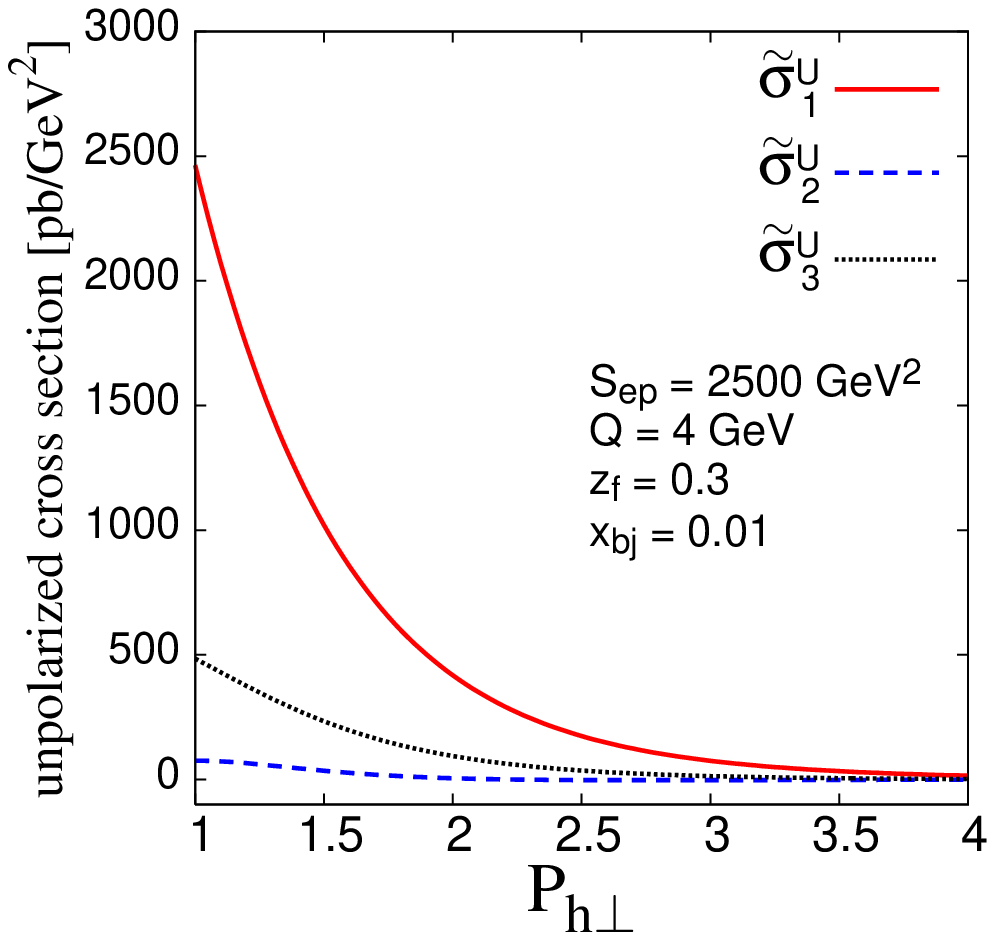}\\
\caption{Same as Fig.~\ref{fig:22}, but for 
EIC kinematics
with $S_{ep}=2500$~GeV$^2$, $Q=4$~GeV, $z_f=0.3$, and $x_{bj}=0.01$.}
\label{fig:32}
\end{figure}
\begin{figure}
\includegraphics[height=7cm]{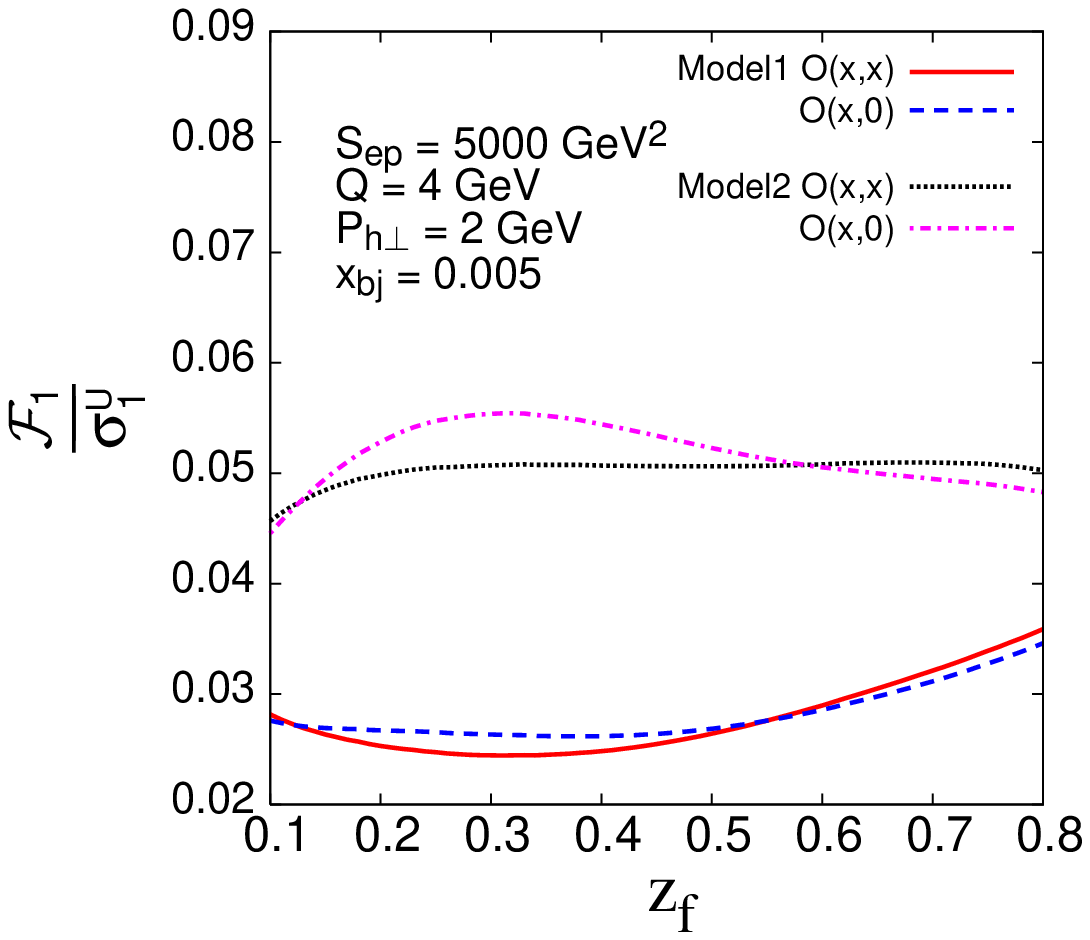}
\hspace{-0.4cm}
\includegraphics[height=7cm]{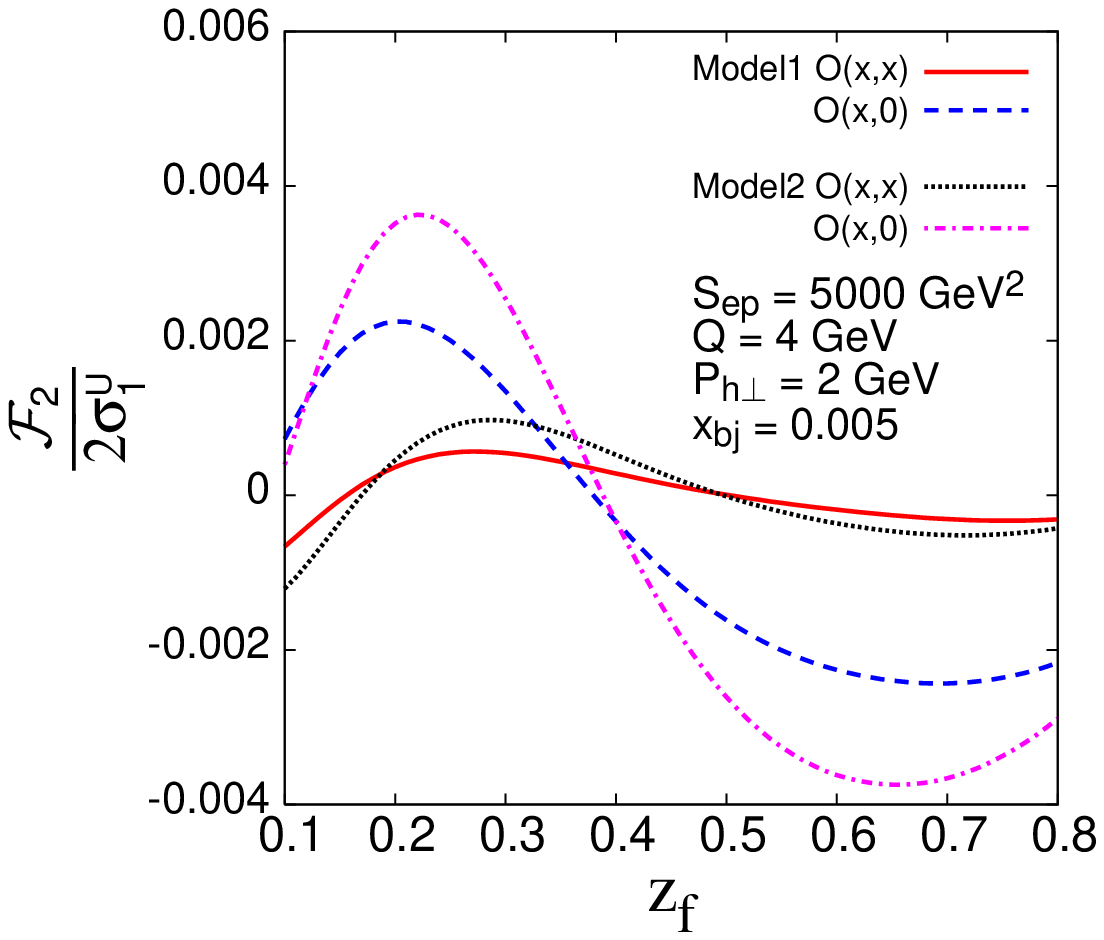}\\
\includegraphics[height=7cm]{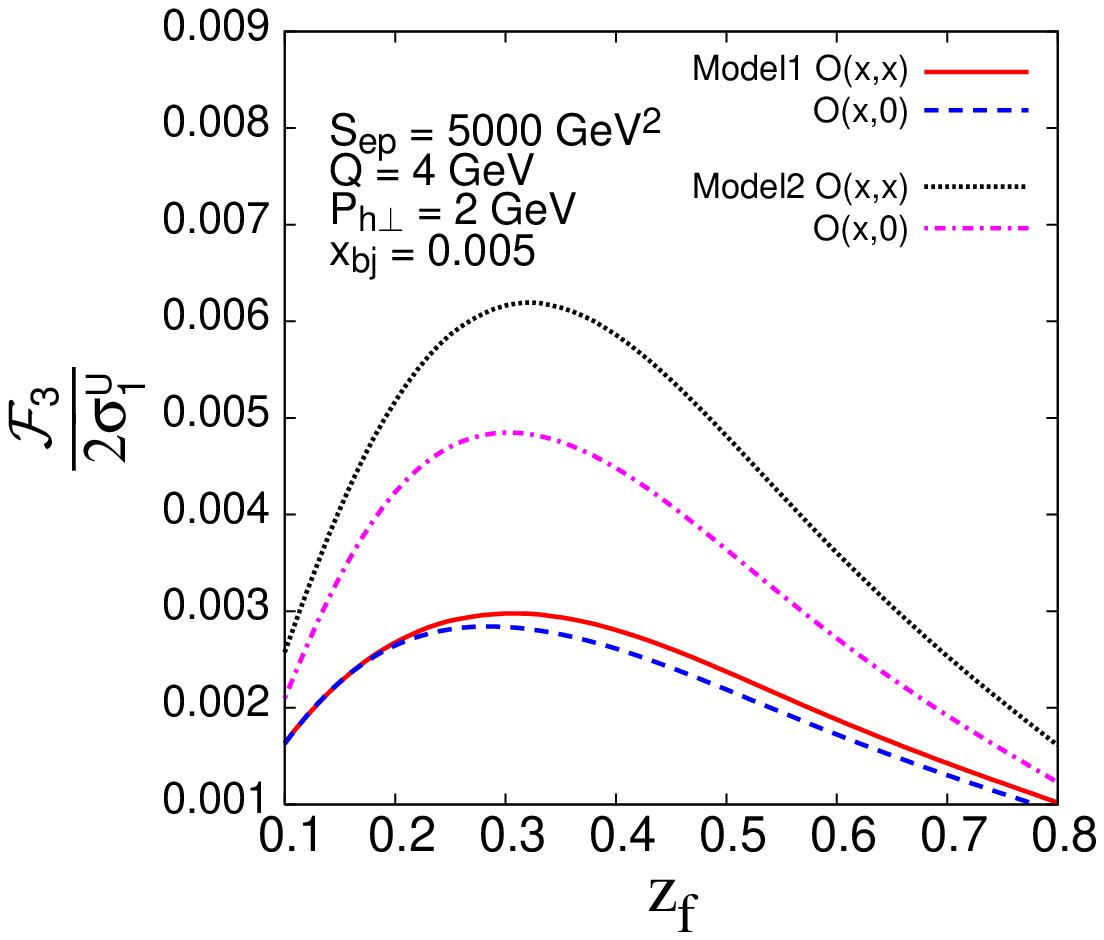}
\hspace{-0.4cm}
\includegraphics[height=7cm]{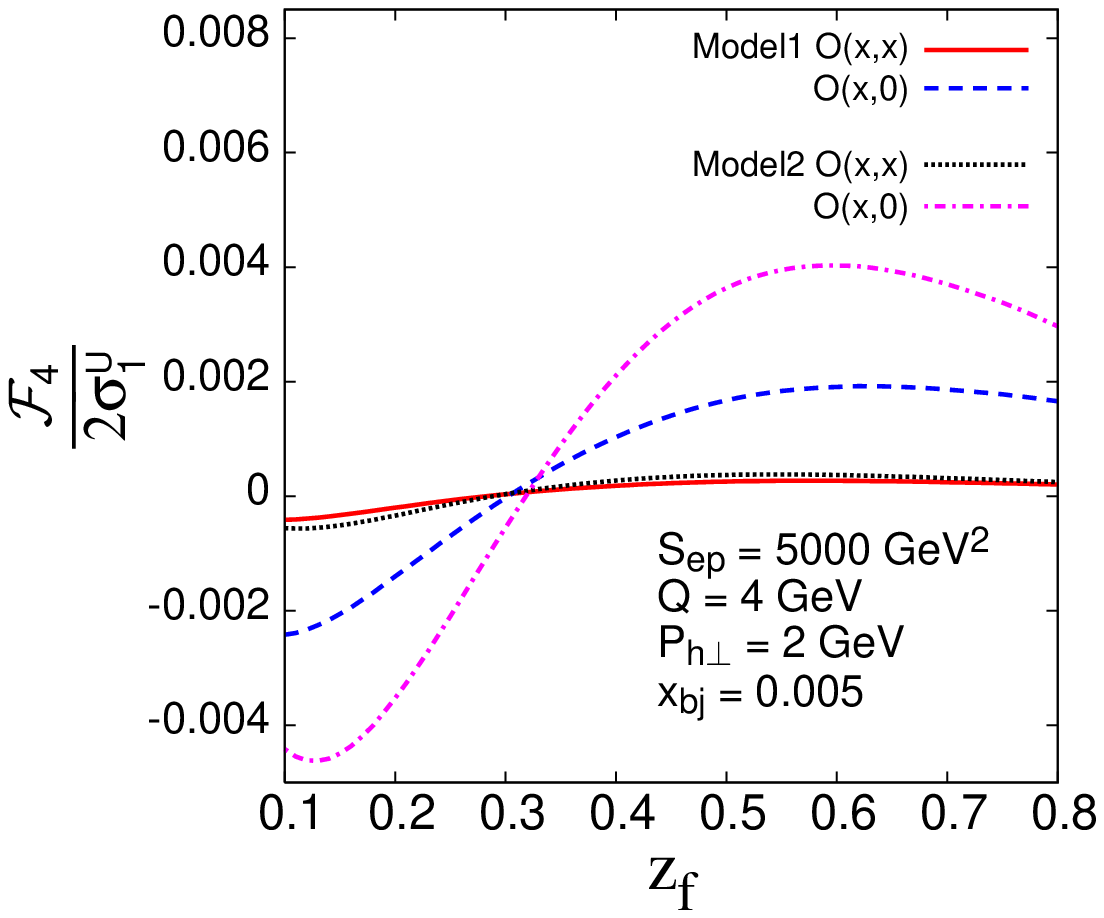}\\
\includegraphics[height=7cm]{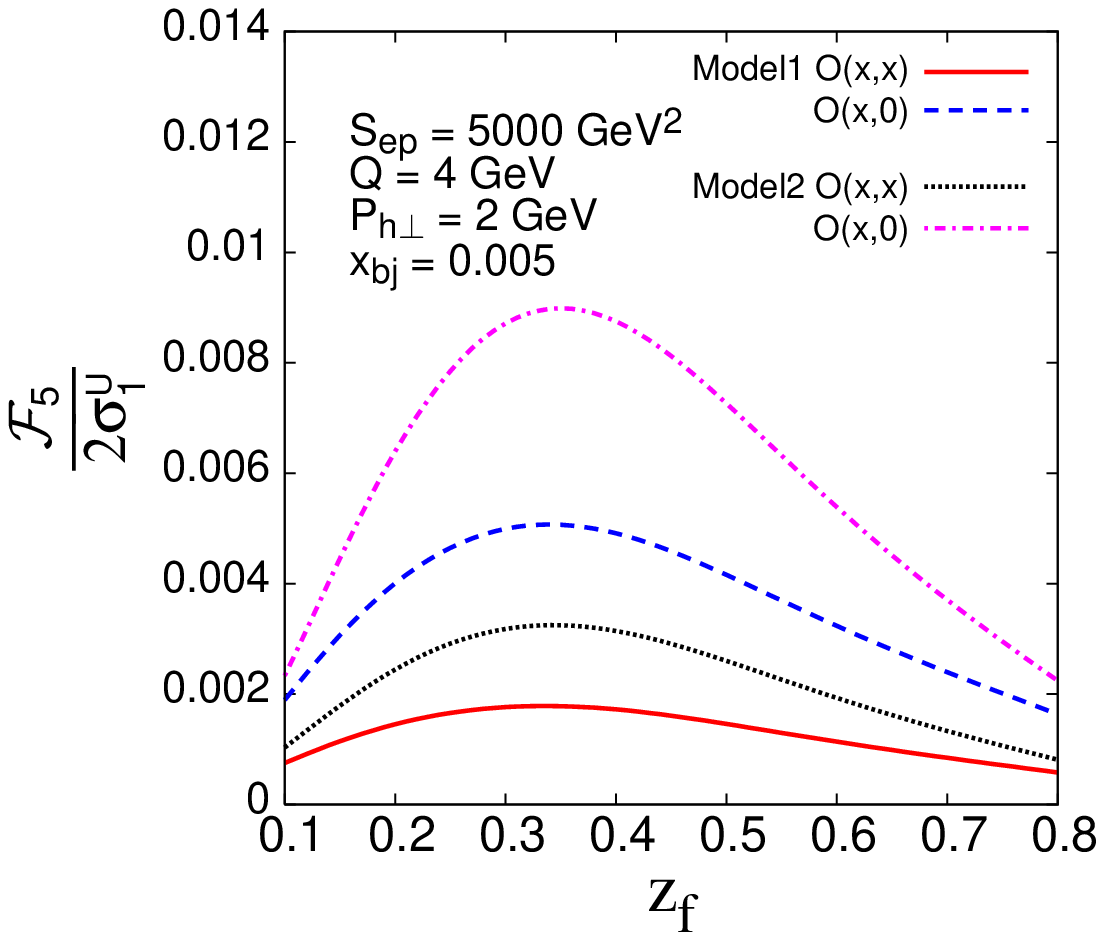}
\hspace{-0.4cm}
\includegraphics[height=7cm]{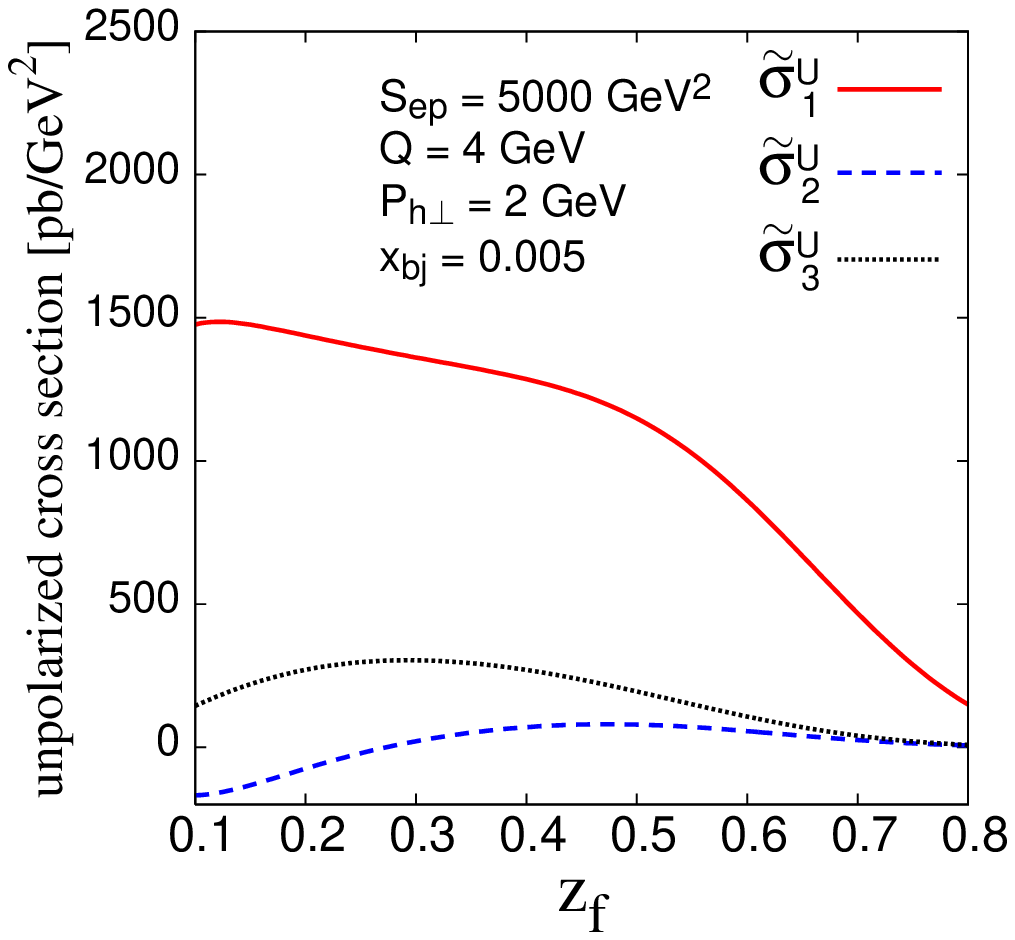}
\caption{Same as Fig.~\ref{fig:21}, but for 
EIC kinematics
with $S_{ep}=5000$~GeV$^2$, $Q=4$~GeV, $P_{h\perp}=2$~GeV, and $x_{bj}=0.005$.}
\label{fig:41}
\end{figure}
\begin{figure}
\includegraphics[height=7cm]{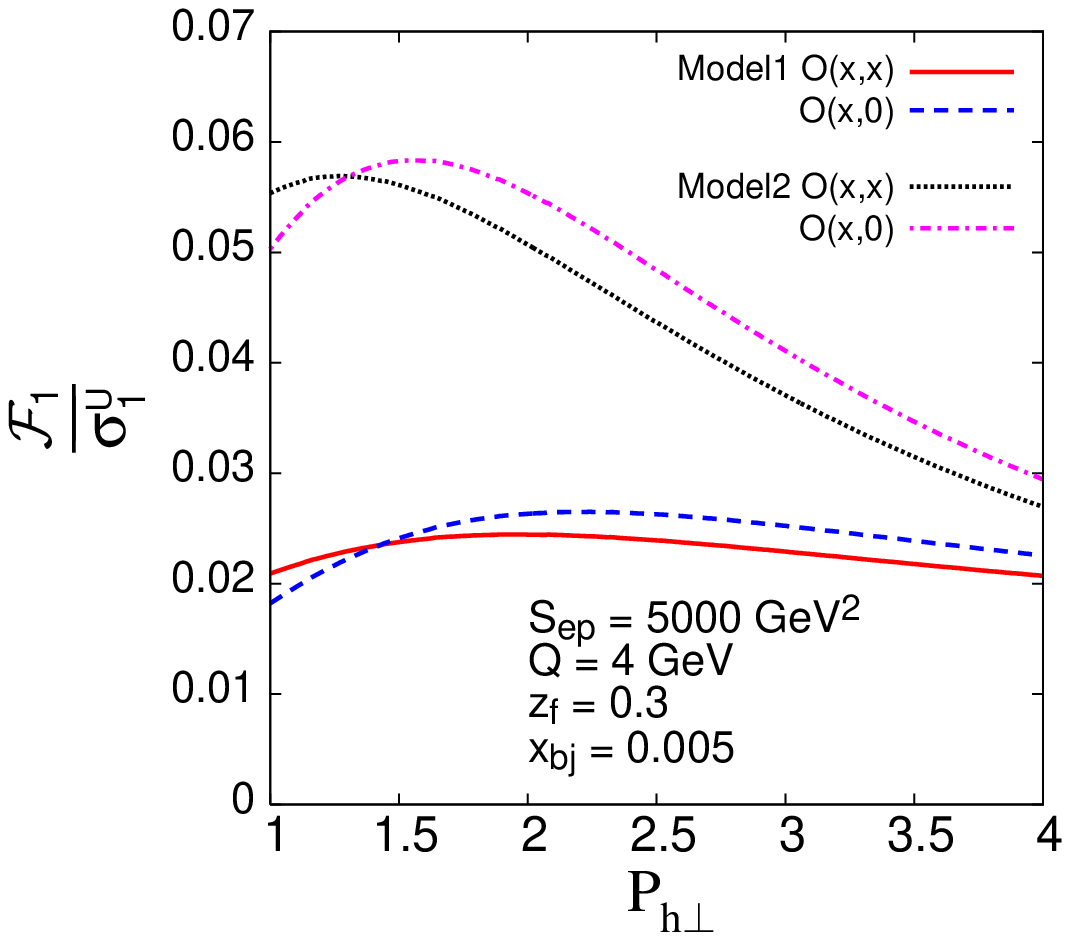}
\hspace{-0.4cm}
\includegraphics[height=7cm]{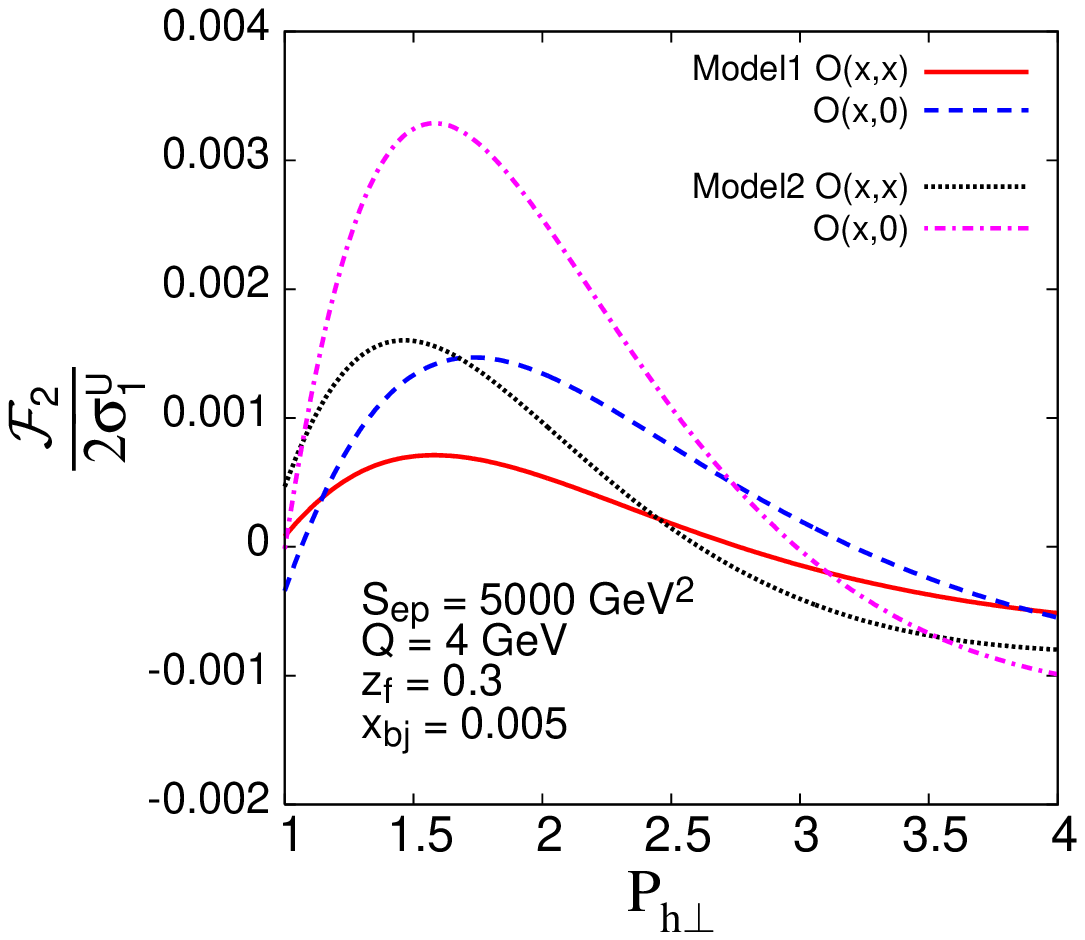}\\
\includegraphics[height=7cm]{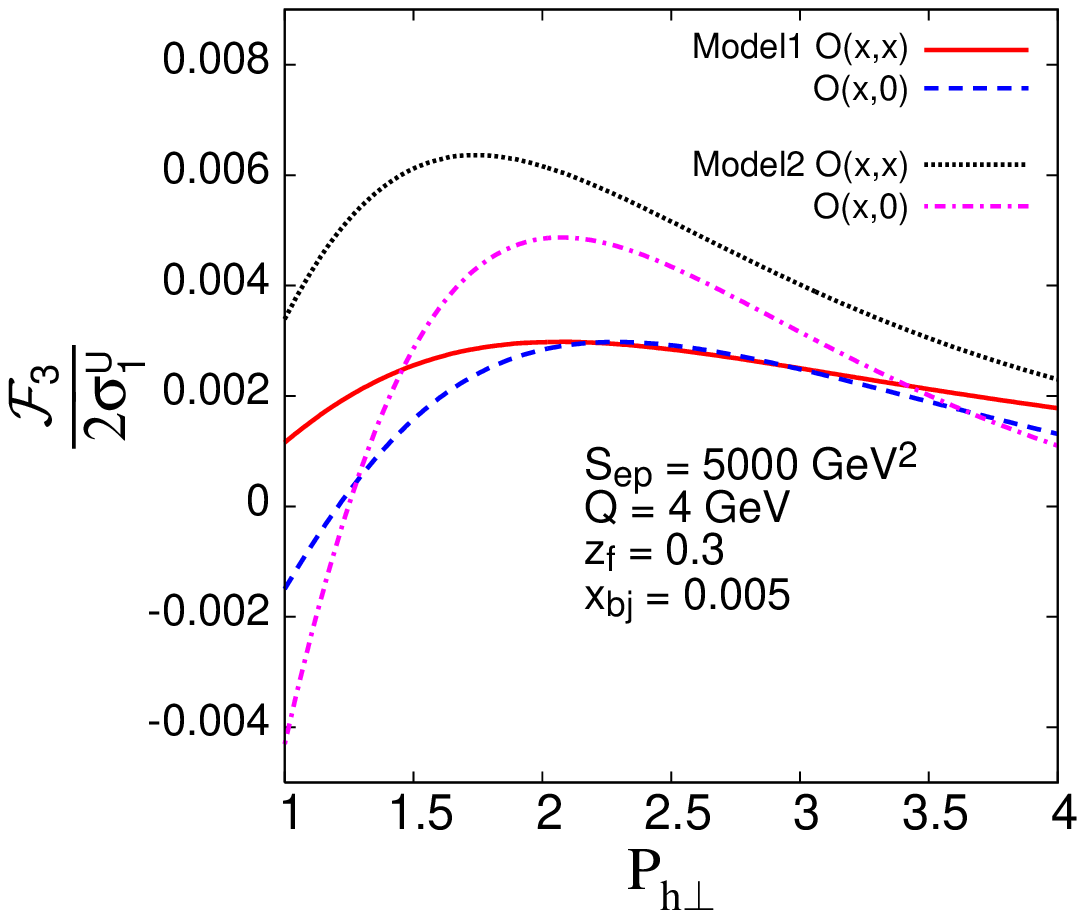}
\hspace{-0.4cm}
\includegraphics[height=7cm]{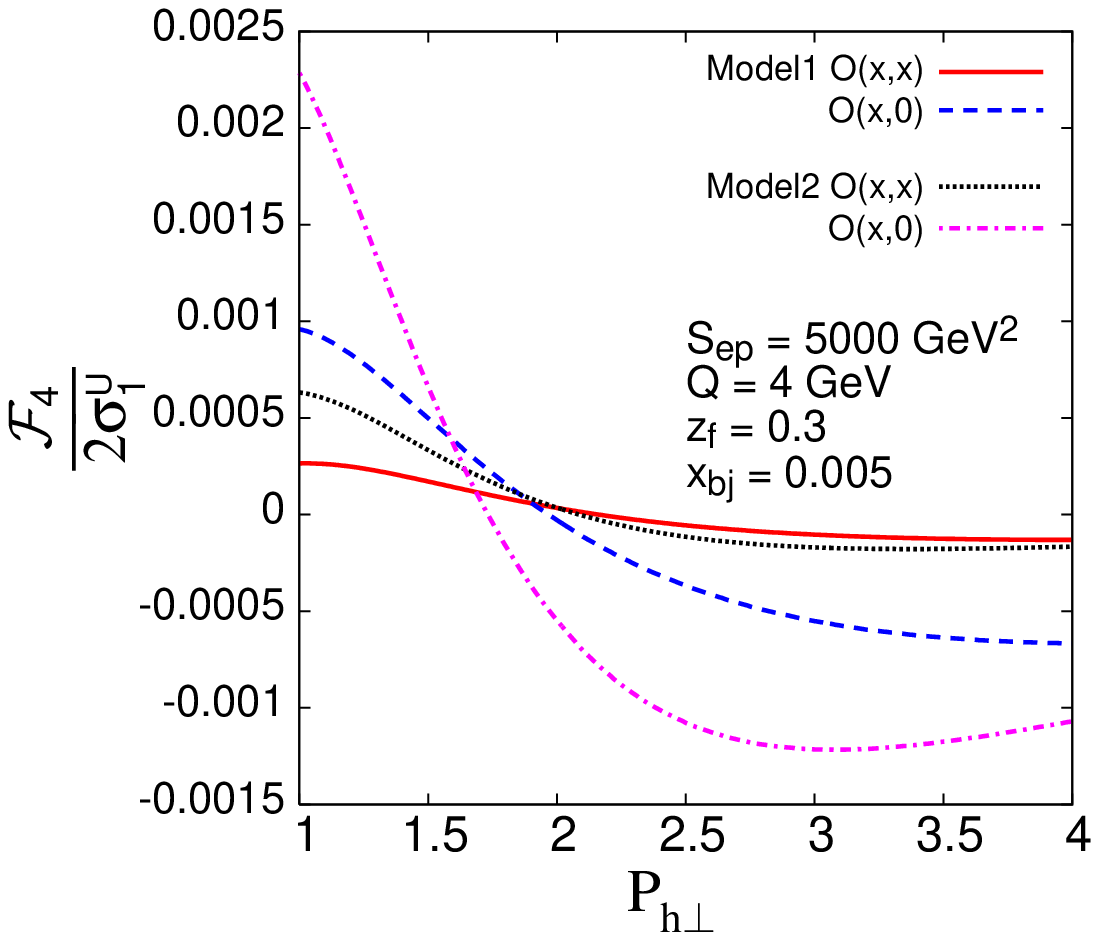}\\
\hspace{-0.4cm}
\includegraphics[height=7cm]{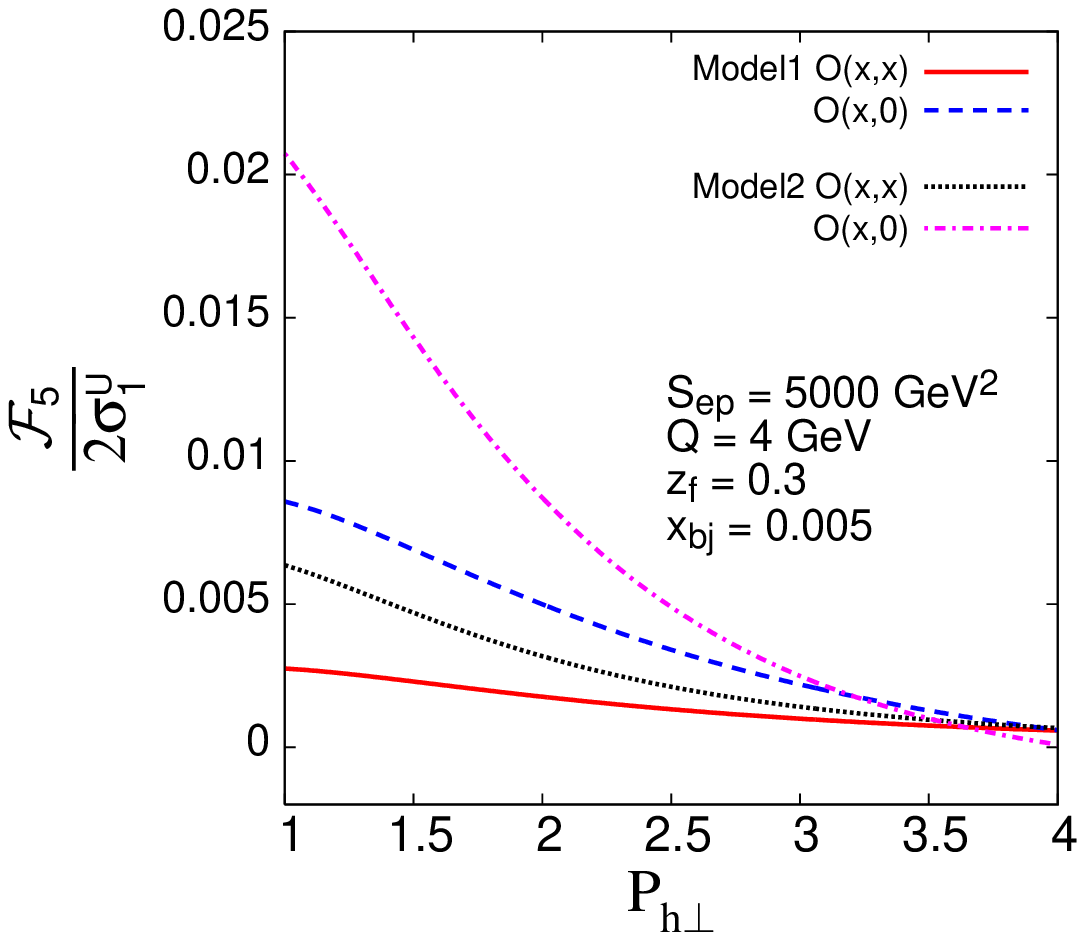}
\hspace{-0.4cm}
\includegraphics[height=7cm]{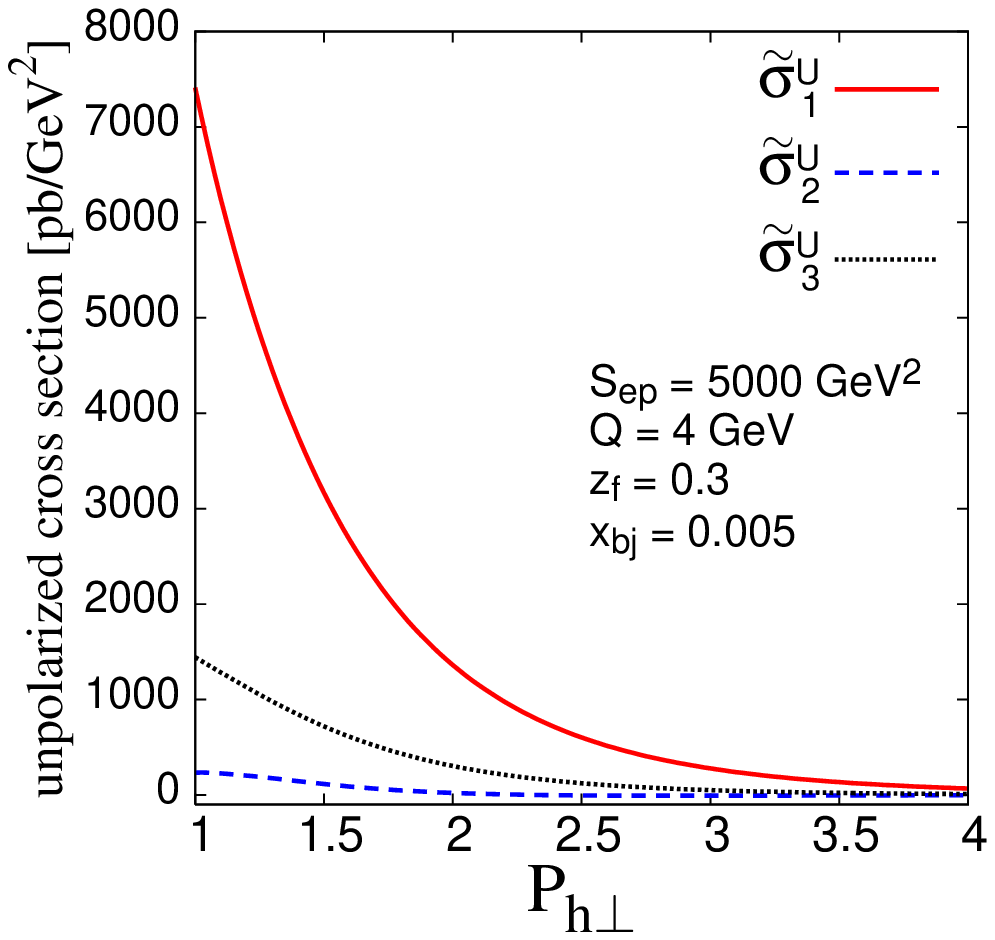}\\
\caption{Same as Fig.~\ref{fig:22}, but for 
EIC kinematics
with $S_{ep}=5000$~GeV$^2$, $Q=4$~GeV, $z_f=0.3$, and $x_{bj}=0.005$.}
\label{fig:42}
\end{figure}

At EIC kinematics
with $S_{ep}=1000$~GeV$^2$, $Q=2.5$~GeV, $P_{h\perp}=2$~GeV, and $x_{bj}=0.01$,
the SSAs in (\ref{ssas})
are, respectively, shown as a function of $z_f$ in 
the first five panels in Fig.~\ref{fig:21},
where 
the contributions 
due to using the models (\ref{case1}) and (\ref{case2}) in (\ref{3gluonresult})
are plotted:
the solid and dashed curves show the contributions 
from $O(x,x)$ and $O(x,0)$ of (\ref{case1}) (Model 1), respectively, to the asymmetries (\ref{ssas}),
so that the difference between those two curves reflect the difference between the 
relevant partonic hard parts
in (\ref{3gluonresult}),
i.e., between $\Delta\hat{\sigma}_k^1$ and $\Delta\hat{\sigma}_k^2$ 
($\Delta\hat{\sigma}_k^3$ and $\Delta\hat{\sigma}_k^4$);
similarly, the dotted and dot-dashed curves show the contributions 
from $O(x,x)$ and $O(x,0)$ of (\ref{case2}) (Model 2), respectively, to the asymmetries (\ref{ssas}).
The behavior of $\sigma_1^U$ arising in the denominator in the SSAs~(\ref{ssas}),
as well as of the other coefficients $\sigma_{2,3}^U$ in the unpolarized cross section~(\ref{unpolresult}),
is shown as $\widetilde{\sigma}_j^U \equiv (2\pi S_{ep} x_{bj}/z_f^2 )\sigma_j^U$
in the last panel in Fig.~\ref{fig:21}
and demonstrates the $D$-meson production rate at EIC.~\footnote{
Changing the variables as $Q^2 \rightarrow y$ ($=p\cdot q/(p\cdot \ell)=Q^2/(x_{bj}S_{ep})$),
$q_T^2 \rightarrow P_{h\perp}^2$,  
and performing the integration over $\chi$ for a fixed $\phi_h$,
(\ref{unpolresult}) becomes
$d^5\sigma^{\rm unpol}/(dx_{bj}dy dz_f dP_{h\perp}^2d\phi_h)=
\widetilde{\sigma}_1^U
+\widetilde{\sigma}_2^U\cos\phi_h
+\widetilde{\sigma}_3^U\cos2\phi_h$.
This choice is convenient for
comparison with the results in \cite{Kang:2008qh}.
The results of $\widetilde{\sigma}_j^U$
shown in the last panel in Figs.~\ref{fig:31}
and \ref{fig:32} are reduced compared with the corresponding numerical results in \cite{Kang:2008qh}.
The reduction mainly comes from the use of the quark mass $m_c=1.5$~GeV
corresponding to the pole mass~\cite{Kneesch:2007ey}, which is larger
compared with the value $m_c=1.3$~GeV used in \cite{Kang:2008qh}.}
We find that the contributions to ${\cal F}_1/\sigma_1^U$ are several percent level and significant,
while those for the other asymmetries are small.
With high energy at EIC,
the $z_f$ dependence of the asymmetries 
is influenced by the different small-$x$ behaviors between the two models (\ref{case1}) and (\ref{case2}).
Similar features are observed also in the $P_{h\perp}$ dependence 
of the SSAs~(\ref{ssas}) and the unpolarized cross section~(\ref{unpolresult}) 
for the fixed  $z_f=0.3$, as shown in  Fig.~\ref{fig:22}.
Here the overall behaviors of the SSAs obey the $1/P_{h\perp}$ falloff 
characteristic of the twist-3 effect.
A remarkable point revealed as a function of $P_{h\perp}$ is
the growth of the asymmetry ${\cal F}_5/(2\sigma_1^U)$,
as well as of the unpolarized cross section~(\ref{unpolresult}),
for decreasing $P_{h\perp}$;
in particular, ${\cal F}_5/(2\sigma_1^U)$ could reach a few percent.
The formulae (\ref{3gluonresult}) for the single-spin-dependent cross section
tells us that the $P_{h\perp}$ as well as $z_f$ dependence of the SSAs measured at EIC provides
the information to determine the $x$ dependence of the gluonic correlation functions
$O(x,x)$, $O(x,0)$, $N(x,x)$, and $N(x,0)$.
Indeed, the different behaviors of the relevant partonic hard parts
in (\ref{3gluonresult}) give rise to very different shape among the five
asymmetries (\ref{ssas}) as a function of $z_f$ as well as of $P_{h\perp}$.
Some asymmetries have maximum and/or minimum, some asymmetries have a node,
and some asymmetries are monotonic functions.

The following figures are same as Figs.~\ref{fig:21}, \ref{fig:22} but for
EIC kinematics with higher energies: 
With $S_{ep}=2500$~GeV$^2$, $Q=4$~GeV, and $x_{bj}=0.01$,
Fig.~\ref{fig:31} (Fig.~\ref{fig:32}) shows the 
results as a function of $z_f$ for $P_{h\perp}=2$~GeV
(as a function of $P_{h\perp}$ for $z_f=0.3$).
With $S_{ep}=5000$~GeV$^2$, $Q=4$~GeV, and $x_{bj}=0.005$,
Fig.~\ref{fig:41} (Fig.~\ref{fig:42}) shows the 
results as a function of $z_f$ for $P_{h\perp}=2$~GeV
(as a function of $P_{h\perp}$ for $z_f=0.3$).
Due to the large values of gluon distribution in the small-$x$ region,
the cross sections are still sizable for those higher energies,
and the features found in the SSAs in Figs.~\ref{fig:21}, \ref{fig:22}
are observed also in Figs.~\ref{fig:31}-\ref{fig:42}, with some of them
being even more pronounced.
We note that all contributions to the asymmetries 
from the functions $O(x,x)$ and $O(x,0)$,
as presented in Figs.~\ref{fig:21}-\ref{fig:42} for the $D$-meson production, 
change their signs for the $\bar{D}$-meson production 
as implied by the factor $\delta_a$ in (\ref{3gluonresult}),
while the contributions from $N(x,x)$ and $N(x,0)$ do not.

The present results with (\ref{case1}), (\ref{case2}) 
indicates that the derivative terms due to $dO(x,x)/dx$ and $dO(x,0)/dx$ in (\ref{3gluonresult})
are the dominant contributions of similar size in the largest 
asymmetry ${\cal F}_1/\sigma_1^U$, to make its value several percent.
On the other hand, for the other asymmetries ${\cal F}_{2,3,4,5}/(2\sigma_1^U)$,
the derivative terms do not give dominant contributions, so that even a nonderivative term
could give the largest contribution; in particular, 
the contribution of the nonderivative term associated with $O(x,0)\Delta \hat{\sigma}_9^4$ 
in (\ref{3gluonresult})
is responsible 
for the above-mentioned growth of ${\cal F}_5/(2\sigma_1^U)$,
which becomes more pronounced for higher energies.

For all cases treated in Figs.~\ref{fig:21}-\ref{fig:42},
$x_{\rm min}$ in (\ref{3gluonresult}) and (\ref{unpolresult})
is given by the formula in the first line in (\ref{xmin}),
so that the value of $x_{\rm min}$ becomes the smallest at $z_f=0.5$.
This property, combined with the fact that the (dominant) derivative-term contributions 
would give ${\cal F}_1/\sigma_1^U \sim 1/(1-x_{\rm min})$,
would explain the existence of the minimum in the solid and dashed curves around $z_f \simeq 0.5$
in the first panel in Fig.~\ref{fig:21}, as discussed in \cite{Kang:2008qh}.
As demonstrated by the dotted and dot-dashed curves in the same figure, however,
such behavior is affected by the small-$x$ behavior of (\ref{case2}) different from
(\ref{case1}), such that the above-mentioned minimum around $z_f \simeq 0.5$ 
could be changed into the maximum.
In Figs.~\ref{fig:31}, \ref{fig:41} with higher energies
the contributions from the small-$x$ region are more important,
and the corresponding minimum or maximum
is less pronounced.

\section{Conclusions}

In this paper we have discussed the SSAs in SIDIS, $ep^\uparrow\to eD X$,
with the $D$ meson having large transverse momentum
through
the photon-gluon fusion mechanism at the twist-3 level,
which is induced by three-gluon correlation inside the nucleon. 
In particular, we have presented a numerical estimate of those SSAs
for the first time with the consistent leading-order accuracy
in QCD, 
using the corresponding collinear factorization formula.  
Gauge invariance and permutation symmetry among the gluons
require the four types of gluonic nonperturbative functions and their derivatives to represent 
the relevant twist-3 mechanism,
as a consequence of the soft-gluon-pole contributions associated with the $3\rightarrow 2$ 
photon-gluon fusion subprocesses or of the master formula with 
the Born-level ($2\rightarrow 2$) 
photon-gluon fusion subprocesses.
The corresponding SSAs receive the five independent azimuthal structures,
and our calculation of them, 
using gluonic nonperturbative functions suggested by the RHIC data 
for $p^\uparrow p \to D X$,
demonstrates good chance to access multi-gluon effects at an 
Electron Ion Collider,
in particular, through the asymmetries with ${\cal F}_1$ and ${\cal F}_5$.
The similar multi-gluon effects also contribute to the SSAs
in $e p^\uparrow \to e \pi X$, as well as in
Drell-Yan and direct-photon productions~\cite{Koike:2011nx}.


\section*{Acknowledgments}
The work of Y.~K. is supported by the Grant-in-Aid for Scientific Research 
No.23540292. 
The work of K.~T. 
is supported in part by the Grant-in-Aid for Scientific Research 
No.23540292 and by
the Grant-in-Aid for Scientific Research on Priority Areas
No.~22011012. 
The work of S.Y. is supported by the Grand-in-Aid for Scientific Research
(No.~22.6032) from the Japan Society of Promotion of Science.





\end{document}